UNIVERSITY OF CALIFORNIA

Los Angeles

# Conformal Invariance in Quantum Field Theory

A dissertation submitted in partial satisfaction of the

requirement for the degree Doctor of Philosophy

in Physics

by

**Abdulaziz Dakhel Alhaidari**

1987

The dissertation of Abdulaziz Dakhel Alhaidari is approved.

______________________________

Robert Blattner

______________________________

V. S. Varadarajan

______________________________

Robert Finkelstein

______________________________

E. T. Tomboulis

______________________________

C. Fronsdal, Committee Chair

University of California, Los Angeles

1987

To my parents, wife and daughters
Amjad and Maram

# TABLE OF CONTENTS

# ACKNOWLEDGEMENTS

I am indebted to C. Fronsdal for all of his advice, guidance and great friendship which made working with him a very pleasant and unforgettable experience. I am also grateful to M. Flato for his continued interest in my work, very stimulating and fruitful discussions. Many thanks go to G. Reyes for the skillful typing of this dissertation. Finally, I acknowledge the support of the Saudi Ministry of Higher Education and the University of Petroleum and Minerals.

# VITA

August 4, 1954  Born, Almadinah, Saudi Arabia

1978  B.S., Electrical Engineering

B.S., Physics

University of Petroleum and Minerals, Dhahran, Saudi Arabia

1978-1980  Teaching Assistant, Department of Physics

University of Petroleum and Minerals, Dhahran, Saudi Arabia

1982  M.S., Physics

University of California, Los Angeles

1983-1987  Research Fellow

Saudi Ministry of Higher Education

1987  Lecturer, Department of Physics

University of California, Los Angeles

# PUBLICATIONS

1. Conformal Invariance and Gauge Fixing in QED (with F. Bayen, M. Flato and C. Fronsdal), Phys. Rev. D **32**, 2673 (1985).
2. Conformal Quantum Yang-Mills, J. Math. Phys. **27**, 2409 (1986).

# ABSTRACT OF THE DISSERTATION

## Conformal Invariance in Quantum Field Theory

by

Abdulaziz Dakhel Alhaidari

Doctor of Philosophy in Physics

University of California, Los Angeles, 1987

Professor Christian Fronsdal, Chair

With the present trend in experimental particle physics of probing yet shorter distances and with the requirement on the theoretical side of renormalizability, conformal invariance becomes an attractive symmetry for particle interactions. That is because conformal field theories form a large class of massless no-scale renormalizable quantum field theories. It is generally believed, however, that renormalization breaks conformal invariance and with the absence of a reported success to find a conformally invariant renormalization scheme, an alternative approach is followed in this thesis. We begin with a conformally invariant theory and study the implications of such idealization.

Conformal field theories are populated with new gauge degrees of freedom and the inclusion, in a natural way, of these auxiliary fields may improve the renormalization program. Another remarkable property is the presence of the vacuum mode in the physical subspace of the gauge potential, believed to be associated with spontaneously generated gauge theories called "zero-center modules."

In spite of all the work that has been done over the years on conformal field theories, conformal electrodynamics has remained only partially developed. New results presented in Part A are as follows: On the classical level, a gauge-fixing condition that is consistent with conformal invariance; in quantum electrodynamics, conformally covariant quantization for massless electrons. In several appendices we treat an alternative method of gauge fixing that involves a complex anomalous conformal degree, electrodynamics in three dimensions, and conformal Liouville theories. In Part B a conformally covariant quantization of non-Abelian gauge theory is presented, and the invariant propagators needed for perturbative calculations are found. The vector potential acquires a richer gauge structure displayed in a larger Gupta-Bleuler triplet whose center is occupied by conformal QED. Path integral formulation and BRST invariance are shown on a formal level in one covariant gauge. Finally in Part C, the gauge structure of the spinor representation of the conformal group is investigated and an attempt to resolve some of the problems present in conformal QED is carried out. Similarities to the singleton program are pointed out and a new reformulation is suggested with the possible inclusion of supersymmetry.

# A. CONFORMAL INVARIANCE AND GAUGE FIXING IN QED

## 1. Introduction

Belief in the benefits of conformal invariance waxes and wanes, fading somewhat after going through a global maximum (so far) in the early seventies. Much current work is at least flirting with conformal invariance; the concept is relevant to the renormalizability of Weinberg-Salam electroweak theory and to gravity [1], to the finiteness of $N = 4$ Yang-Mills theory [2], to the problem of triviality of $\lambda\phi^4$ field theory in four dimensions [3] and to some of the Grand Unification Theories. But a real understanding of the connection between conformal invariance and renormalizability is lacking. The simplest example worthy of serious study would be conformal QED, and this theory does not exist yet. In this section of the thesis, we shall come very close to a complete formulation of conformal QED; close enough to foresee a full development of conformal renormalization in the very near future.

Conformal invariance of electrodynamics began in 1909 with Cunningham and Bateman [4]. Group theoretical aspects were studied during the sixties [5] and the problem of conformally invariant gauge fixing was first investigated at about the same time [6,7]. Baker and Johnson [8] initiated the study of conformal invariance as a computational tool and this aspect was taken up by others after Migdal [9] suggested the conformal bootstrap. Unfortunately, these pioneers lost the direction and applied themselves to "Euclidean electrodynamics" instead [10]. See also the work of Fradkin and Palcik [11].

Manifest conformal invariance was invented by Dirac already in 1936 [12], and received major developments in the hands of Mack and Salam [13]. These papers do not carry any quantization, even on a formal level. Adler [14] went further, developing a Feynman calculus in Dirac's positive 6-cone notation, but his photon propagator is not covariant and there is no field quantization in his paper. The curious absence of field quantization in all the papers that have been quoted so far can be understood if one remembers that the motivation for much of this work was the hope for rationalizing computations in a theory that was seen as being already in its ultimate formulation. The new viewpoint [15] is quite different; we want to know precisely how conformal invariance affects the renormalization program. This attitude does not conflict with that of Callan, Coleman and Jackiw [16], who studied the inverse problem, the effect of renormalization on conformal invariance; it is complementary to it.

Of course, conformal invariance is incompatible with the boundary conditions needed to define an S-matrix, and for precisely this reason it is crucial to push it as far as possible. Recently it was shown [15] that free quantum fields (electromagnetic potentials) can be defined that satisfy conformally invariant commutation relations, at the cost of introducing an extra scalar field that provides the necessary canonical momenta, conjugate to the gauge modes. The fact that there are more gauge modes in a conformally invariant theory was recognized already by Dirac [12]. In later work it takes the form of a relaxation of the gauge-fixing condition; the Lorentz condition $\partial \cdot A = 0$ is replaced by the conformally invariant $\Box \partial \cdot A = 0$ [7].

A related phenomenon occurs in the conformally invariant treatment of the Dirac electron field - and, in fact, also in scalar conformal electrodynamics. Unphysical field components appear, and the fact that they couple only to a (generalized) free field is only partly reassuring. This problem first turned up in Dirac's paper [12]; Mack and Salam [13] made it a question of principle to decouple the unphysical field - at least in one electromagnetic gauge. Reference 15 did not deal adequately with this. A fairly complete treatment will be presented here, for both scalar and spinor conformal QED.

An attempt at conformal quantization of the massless electron field by Adler [14] led to the discovery of an incompatibility between the field equations and the commutation relations: no conformally invariant two-point function could be found that satisfied the free field equations. This problem is treated here; it turns out that what one needs is a slight generalization of the concept of conformal degree.

It is generally believed that the electromagnetic potential retains its normal (geometrical) degree, but that some anomaly appears in the conformal degree of the electron field $\psi$. In Appendix A we have developed an alternative treatment of conformal gauge fixing, in which $\psi$ is given a complex conformal degree, the real part of which is just the normal degree. The following remarks apply to the real part of the conformal degree of the electron field, without any prejudice to the possibility of a non-vanishing imaginary part. An abnormal conformal degree may of course accompany the introduction of an electron mass, though the very concept of conformal degree stands to loose its meaning when conformal invariance is abandoned. The question that may be discussed is whether it will be necessary to assign an anomalous degree to the massless electron field. Perhaps the most enduring contribution of this work is the discovery that the electron field, in conformally invariant quantum field theory, is a gauge field. Conformally invariant quantization requires a full Gupta-Bleuler, indefinite-metric quantization scheme for the electron field, just as it does for the photon field. This is, of course, also true in supersymmetric field theories, if the electron field belongs to the gauge multiplet. One must therefore conclude that, if the arguments that have been advanced [17] to show that the photon field cannot develop an anomalous degree are correct, then they should apply with equal force to the massless electron field. We want to point out, however, that a degree of skepticism toward these same arguments is justified. The main point is that current conservation is conformally invariant only if the degree of the current is exactly 3. But if the degree of the electromagnetic potential is $1+\varepsilon$, with $\varepsilon > 0$, then the associated representation of the conformal group is fully reducible, and current conservation is no longer the only way to make the theory unitary. So we cannot exclude the possibility that all the fields may develop an anomalous conformal degree. Another possibility, already suggested by the classical theory, is that logarithms may appear in the unphysical part of the free propagators. We develop one of these options in the case of scalar QED and explore the other in dealing with spinor electrodynamics. Both alternatives are available in either case.

## 2. Classical Conformal Gauge Fixing

The free Maxwell equations in Minkowski space,

$$\partial_\mu F_{\mu\nu} = 0, \qquad F_{\mu\nu} \equiv \partial_\mu A_\nu - \partial_\nu A_\mu,$$

are invariant under the transformations of the conformal group $\mathscr{C}$. [Strictly, Minkowski space must be compactified.] Looking for conformally invariant gauge-fixing conditions, one finds essentially only two possibilities [7]:

$$\partial_\mu(A^2 A_\mu) = 0, \qquad \Box \partial \cdot A = 0.$$

We have nothing to say about the nonlinear equation and consider only the pair

$$\partial_\mu F_{\mu\nu} = 0, \tag{2.1}$$

$$\Box \partial \cdot A = 0. \tag{2.2}$$

Actually, the second equation is not invariant by itself, but the pair is. If

$$I_\mu(x) \equiv \partial_\nu F_{\nu\mu}(x), \qquad G(x) \equiv \Box \partial \cdot A(x),$$

then the effect of the special conformal transformation

$$x \to x' = \omega(a,x)^{-1}(x + x^2 a), \qquad \omega(a,x) \equiv 1 + 2a \cdot x + a^2 x^2, \qquad a \in R^4, \tag{2.3}$$

on $(I_\mu, G)$ is as follows:

$$I'_\mu(x') = \omega(a,x)^4 \left(\partial x'_\mu / \partial x_\nu\right) I_\nu(x),$$

$$G'(x') = \omega(a,x)^4 \left\{ G(x) - 2\partial_\mu \ln|\omega| I_\mu(x) \right\}.$$

This is a non-decomposable representation, an invariant subspace being defined by Maxwell's equations $I_\mu(x) = 0$.

The systems (2.1) and (2.2) can be derived from an action principle with the Lagrangian density

$$\mathscr{L}_M = -\frac{1}{4} F_{\mu\nu} F_{\mu\nu} - \frac{1}{2} Z \Box \partial \cdot A, \tag{2.4}$$

where $Z$ is a scalar field. More precisely, one obtains the equations

$$\partial_\mu F_{\mu\nu} + \frac{1}{2} \Box \partial_\nu Z = 0, \tag{2.5}$$

$$\Box \partial \cdot A = 0. \tag{2.6}$$

This is conformally invariant, provided that $(A_\mu, Z)$ defines another non-decomposable representation. Under special conformal transformations

–3–

$$A'_\mu(x') = \omega(a,x)^2 \left(\partial x'_\mu / \partial x_\nu\right)\left[A_\nu(x) - \partial_\nu \ln|\omega| Z(x)\right], \tag{2.7}$$

$$Z'(x') = Z(x). \tag{2.8}$$

The Lagrangian (2.4) and the transformation laws (2.7) and (2.8) were proposed in Ref. 7 and also by Zaikov [17].

Next, consider the coupled Maxwell-Dirac system (with vanishing electron mass)

$$\partial_\mu F_{\mu\nu} = J_\nu, \tag{2.9}$$

$$(\not{\partial} - ie\not{A})\psi = 0, \tag{2.10}$$

$$\Box \partial \cdot A + 4\rho = 0. \tag{2.11}$$

Here $J_\mu = -e\bar{\psi}\gamma_\mu\psi$ and the quantity $\rho$ (a scalar field) has been included in the gauge condition in recognition of the fact that the system is not conformally covariant in the absence of such a corrective term. The action of special conformal transformations on $\psi$ is given by

$$\psi'(x') = \omega(a,x)(1 + \not{a}\not{x})\psi(x). \tag{2.12}$$

*Proposition.* The system of Eqs. (2.9)-(2.11) is covariant under the transformations of the conformal group if and only if the field $\rho$ cotransforms in the following manner (under special conformal transformations):

$$\rho'(x') = \omega(a,x)^4\left\{\rho(x) + \tfrac{1}{2}J_\mu(x)\partial_\mu \ln|\omega|\right\}. \tag{2.13}$$

This means that $\rho$ is determined by $J_\mu$ up to a conformal scalar of degree $-4$. In particular, this condition is satisfied if

$$\rho = (1/6p) J_\mu \partial_\mu \ln|P(\bar{\psi}\psi, \bar{\psi}\gamma_5\psi)|, \tag{2.14}$$

where $P(a,b)$ is a real, homogeneous polynomial of total degree $p > 0$ in $a$ and $b$. Gauge fixing of the form (2.11) and (2.14) has been proposed by Sotkov and Stoyanov [18].

The Lagrangian density

$$\mathcal{L} = -\frac{1}{4} F_{\mu\nu} F_{\mu\nu} - 2Z\left(\rho + \tfrac{1}{4}\Box\partial\cdot A\right) - J\cdot A + \frac{i}{2}\left(\bar{\psi}\not{\partial}\psi - \not{\partial}\bar{\psi}\,\psi\right). \tag{2.15}$$

gives the following Euler-Lagrange equations:

$$\partial_\mu F_{\mu\nu} + \tfrac{1}{2}\Box\partial_\nu Z = J_\nu, \tag{2.16}$$

$$\Box\partial\cdot A + 4\rho = 0, \tag{2.17}$$

$$(\not{\partial} - ie\not{A})\psi + 2i\frac{\delta(Z\rho)}{\delta\bar{\psi}} = 0. \tag{2.18}$$

Under the stated conditions on $\rho$, the Dirac current is conserved, and it follows that $Z$ is a generalized free field,

$$\Box^2 Z = 0. \tag{2.19}$$

We conclude that conformally invariant gauge fixing is an available option in classical, massless electromagnetic theory. The Lagrange multiplier $Z$ can be taken to vanish after it has served to derive the equations of motion. [If $Z$ vanishes on the initial data, then it vanishes at all times by Eq. (2.19).] The logical basis, and even naturalness of this prescription, will become much clearer after quantization. An alternative treatment of gauge fixing is given in Appendix A.

## 3. Conformal Electromagnetic Field Quantization

In Dirac's six-cone notation [12,13] the action for the electromagnetic potential in the presence of an external current is [15]

$$\int (dy)\left(\tfrac{1}{2} a_\alpha \partial^2 a_\alpha - a \cdot j\right). \tag{3.1}$$

The fields are defined on the projective cone $y^2 = 0$, $\lambda y \simeq y$ for $\lambda \neq 0$, and the indices range from 0 to 5. The operator $\partial^2$ is the d'Alembertian defined by the pseudo-orthogonal metric $\delta = \mathrm{diag}(+----+)$.

It has been common practice to impose the condition $y \cdot a = 0$, but it was shown in Ref. 15 that this is incompatible with quantization. In fact,

$$y \cdot a = 0, \quad \text{is the Lorentz condition} \tag{3.2}$$

of conformal QED. We are therefore forced to accept an extra scalar field; this is the field that is denoted $A_+$ in Minkowski notation and that turns out to be closely related to the Lagrange multiplier $Z$ of Sec. 2. At the end of this section we shall give a simple account of the reason why this extra field is needed in field quantization.

The quantum field operator is defined by

$$\hat{a}_\alpha(y) = \sum_i a_\alpha^i(y) b_i + \text{H.c.},$$

where $b_i$, $b_i^*$ is a set of annihilation, creation operators satisfying canonical commutation relations, and $\{a_\alpha^i\}$ is a set of solutions of the free field equation $\partial^2 a_\alpha(y) = 0$. The linear span of these solutions carries the following nondecomposable (Gupta-Bleuler triplet) of $\mathscr{C}$ (Ref. 15):

$$D\left(1,\tfrac{1}{2},\tfrac{1}{2}\right) \to \left\{\begin{array}{c} D(2,1,0) \oplus D(2,0,1) \\ \oplus \mathrm{Id} \end{array}\right\} \to D\left(1,\tfrac{1}{2},\tfrac{1}{2}\right)$$

The physical photon states are in the central subquotient, the scalar multiplet on the left is precisely the representation associated with the dipole ghost $\Box^2 Z = 0$ [Eq. (2.19)]. The invariant two-point function (the reproducing kernel for the above representation) is, simply,

$$\langle \hat{a}_\alpha(y)\hat{a}_\beta(y')\rangle = \delta_{\alpha\beta}(y\cdot y')^{-1}, \tag{3.3}$$

where the distribution on the right-hand side is defined as the unique invariant, homogeneous generalized function with no negative-energy Fourier coefficients. Reference 15 may be consulted for details.

In Minkowski notation the six-vector $\{a_\alpha\}$ is replaced by a vector field $\{A_\mu\}$ and scalar fields $A_+$, $A_B$. The current also has two extra components, and the action is

$$I[A] - \int d^4x \left(A\cdot J + 2A_+ J_B + 2A_B J_+\right), \tag{3.4}$$

$$I[A] = -\int d^4x \left(\tfrac{1}{2}\partial_\mu A_\nu \partial_\mu A_\nu + 4A_B \partial\cdot A + 2\partial_\mu A_+ \partial_\mu A_B + 8A_B A_B\right). \tag{3.5}$$

The external current $J_+$, $J_\mu$, $J_B$ comes from $\{j_\alpha\}$, $\alpha = 0,1,..,5$. The infinitesimal transformations of $A_+$, $A_\mu$, $A_B$ were given in Ref. 15, and the finite transformations were given by Todorov [18]. It will soon be seen that $A_+$ can be identified with the Lagrange multiplier Z of the preceding section, and that the transformations of $\{A_+, A_\mu\}$ agree with (2.7) and (2.8).

The field equations are

$$\Box A_\mu + 4\partial_\mu A_B = J_\mu, \tag{3.6}$$

$$\Box A_+ - 2\partial\cdot A - 8A_B = J_+, \tag{3.7}$$

$$\Box A_B = J_B. \tag{3.8}$$

The conformal degrees of $A_+$, $A_\mu$, $A_B$ are 0,1,2, and those of $J_+$, $J_\mu$, $J_B$ are 2,3,4. A gauge transformation is a transformation of the type

$$\left(A_+, A_\mu, A_B\right) \to \left(A_+, A_\mu + \partial_\mu \Lambda, A_B + \tfrac{1}{4}\Box\Lambda\right). \tag{3.9}$$

The Lagrangian and the field equations are invariant, in the free case, under the restricted gauge group defined by $\Box^2 \Lambda = 0$. The interaction Lagrangian is invariant under the unrestricted gauge group, provided the external current satisfies the modified conservation law [15]

$$\partial\cdot J + \tfrac{1}{2}\Box J_+ = 0. \tag{3.10}$$

In view of the field equations, this is equivalent to the statement that $A_+$ is a (generalized) free field:

$$\Box^2 A_+ = 0. \tag{3.11}$$

The Lorentz condition (3.2) is expressed by $A_+ = 0$, but of course this cannot be imposed on the quantum field operator. Just as the conventional Lorentz condition of QED, it must be replaced by an initial condition on the physical states; Eq. (3.11) guarantees that the condition is preserved by the time development.

There are models of matter in which $J_+$ vanishes. Setting $J_+ = 0$ and eliminating $A_B$ one reduces the system (3.6)-(3.8) to

$$\partial_\mu F_{\mu\nu} + + \tfrac{1}{2}\Box\partial_\nu A_+ = J_\nu, \tag{3.12}$$

$$\tfrac{1}{2}\Box^2 A_+ - \Box \partial \cdot A = 4 J_B. \tag{3.13}$$

These are the equations proposed by Zaikov (Ref. 17), $A_+$ being his $R$ field and $J_B$ his $D$ current. If the current is conserved they simplify

$$\partial_\mu F_{\mu\nu} + \tfrac{1}{2}\Box\partial_\nu A_+ = J_\nu, \tag{3.14}$$

$$\Box \partial \cdot A + 4 J_B = 0. \tag{3.15}$$

These equations agree with Eqs. (2.16) and (2.17), if $A_+ = Z$ and $\rho = J_B$. This confrontation between the classical theory and the would-be quantum field theory is important for both, for the following reason:

On the one hand, the *ad hoc* treatment of the Lagrange multiplier $Z$ in the classical theory is completely justified by its role as Lorentzian constraint field in quantum field theory. On the other hand, we learn something about the mysterious quantity $J_B$: its role is the role of $\rho$ in the classical theory, serving to make the gauge condition conformally invariant. We also find out that it may be possible to express $J_B$ entirely in terms of physical matter fields. In fact, it is interesting to recover Eq. (2.14) from the present vantage point.

Let *s* be any scalar field, then the following equation (in Dirac notation) is conformally covariant:

$$j_\alpha \partial_\alpha s(y) = 0.$$

(This is meaningful if $y \cdot j = 0$; that is, if $J_+ = 0$.) Convert to Minkowski notation,

$$J_\mu \partial_\mu S + 2 J_B \partial_+ S = 0.$$

If *s* is homogeneous of degree $q$, then $\partial_+ S = qS$, and thus

$$J_B = -\frac{1}{2q} J_\mu \partial_\mu S.$$

This is in perfect agreement with Eq. (2.14), if we identify $J_B$ in Eq. (3.15) with $\rho$ in Eq. (2.11), and recall that the degree of $\psi$ is $\frac{3}{2}$ so that the degree of $P$ is $3p$. We conclude, therefore, that a fairly well-defined expression for $J_B$, that respects conformal covariance, can be had by taking $s = \bar{\psi}\psi$ or some other homogeneous polynomial in the physical matter fields. Another possibility, suggested by Todorov [18], is to take $L = \ln S$ as a new and independent field that, instead of being homogeneous, satisfies the condition $y \cdot \partial L = -1$. Then the logarithm can be avoided, but $L$ becomes a dipole ghost.

Equation (3.2), in Minkowski notation, reads

$$A_+(x) = 0 \text{ (Lorentz condition)}.$$

We end this section with a direct explanation of why this condition cannot be imposed on the field operators without loss of conformal invariance. Suppose first that the field operator $A_\mu$ transforms in the usual way under the conformal group, and that, in particular, the generators $K_\mu$ of special conformal transformations act as follows (correct if $A_+ = 0$)

$$K_\mu A_\nu = \left[x^2 \partial_\mu - 2x_\mu (x \cdot \partial + 1)\right] A_\nu + 2\left(x_\nu A_\mu - \delta_{\mu\nu} x \cdot A\right).$$

The two-point function

$$\langle A_\mu(x) A_\nu(x') \rangle = -2\delta_{\mu\nu} \left[(x-x')^2\right]^{-1}. \tag{3.16}$$

is then formally covariant. However, in order to define the free quantum field operator it is necessary to include gauge modes and to choose the gauge. Equation (3.16) holds true in the Feynman gauge, in which the gauge modes satisfy $\partial \cdot A = 0$ ($A_\mu = \partial_\mu \phi$ with $\Box \phi = 0$), but this gauge is not conformally invariant. The conformally invariant gauge condition $\Box \partial \cdot A = 0$ is weaker and the conformally covariant field operator thus contains more gauge modes ($A_\mu = \partial_\mu \phi$ with $\Box^2 \phi = 0$). The field $A_+$ is needed to provide the extra gauge modes with canonically conjugate field modes.

In the free field case, if $A_+ = 0$, then the field $A_B$ is just $\partial \cdot A$ and evidently contains the extra gauge modes. If $A_+ \neq 0$ the relationship between $A_B$ and $\partial \cdot A$ is given by Eq. (3.7). The complete set of two-point functions is given by Eq. (3.16) and

$$\langle A_\mu(x) A_+(x') \rangle = 2(x_\mu - x'_\mu)\left[(x-x')^2\right]^{-1},$$

$$\langle A_+(x) A_B(x') \rangle = -\left[(x-x')^2\right]^{-1},$$

$$\langle A_+(x) A_+(x') \rangle = 1.$$

[The whole set is equivalent to Eq. (3.3).] It is clear that $A_+$ is conjugate to $A_B$. The action of $K_\mu$ on $A_+$, $A_\mu$, $A_B$ was given in Ref. 15.

## 4. Conformal Scalar QED

The problem consists of the fact that the formal, gauge-invariant, minimal coupling introduces a current component $J_B$ that turns out to involve unphysical fields. For example:

Spinor conformal QED, in the manifestly covariant formalism [12,13], deals with an eight-spinor representation of $\mathscr{E}$ that is non-decomposable. The physical representation is the quotient. The free Lagrangian is "intrinsic," which means that the unphysical fields of the invariant subspace do not contribute. But the interaction is not intrinsic, and this precisely because $J_B$ involves the unphysical components.

Scalar conformal QED, in the manifestly covariant formalism, is the theory of a complex, scalar field $\phi$ of degree of homogeneity $-1$ on the projective six-cone. The field satisfies $\partial^2 \phi(y) = 0$, and "manifest minimal coupling" is the modification

$$(\partial_\alpha - iea_\alpha)^2 \phi(y) = 0.$$

But this makes no sense, for $\partial_\alpha \phi$ and $\partial_\alpha a_\beta$ are undefined for fields on the cone. Since $y^2 = 0$, $\phi = \phi' + y^2 \sigma$ means the same as $\phi = \phi'$, and $\sigma$ is "unphysical." But differentiation gives $\partial_\alpha \phi = \partial_\alpha \phi' + 2 y_\alpha \sigma$ at $y^2 = 0$, so $\partial_\alpha \phi$ has unphysical components of the form $y_\alpha \sigma$. These form an invariant subspace of the module $\{\partial_\alpha \phi\}$, so the situation is basically the same as in the spinor case. Actually, it is worse in the scalar case since $\partial_\alpha a_\beta$ also intervenes.

We approach these problems in the following spirit. Whatever one thinks about the unphysical field components, it is self-evident that they have to be quantized. Therefore, let us carry out the quantization in detail and examine the difficulties afterward.

Minkowski space is the projective six-cone $y^2 = 0$ in $R^6$, and the physical content of the scalar field on $R^6$ is determined by the values that it takes on the cone. However, since the interaction depends, at least formally, on the normal derivative, we must retain the term of order $y^2$ in the expansion

$$\phi(y) = x_+^{-1} \left[ \Phi(x) + x_B \sigma(x) + O(x_B^2) \right], \tag{4.1}$$

$$x_+ = y_4 + y_5, \qquad x_B = y^2/(y_4 + y_5)^2. \tag{4.2}$$

Here $\Phi$ is the physical scalar field on Minkowski space. The unphysical field $\sigma$ does not appear in the free Lagrangian, and in the interaction Lagrangian it contributes only to the unphysical component $J_B$ of the current. (It is possible to include, in the free Lagrangian, a term $\Phi^* \sigma + \Phi \sigma^*$, without destroying the conformal invariance of the action.)

The quantum field operator

$$\hat{\phi}(y) = \sum_i \left[ \phi^i(y) c_i + \bar{\phi}^i(y) d_i^* \right]. \tag{4.3}$$

is constructed from positive-energy solutions $\phi^i$ of the free field equation, with canonical annihilation, creation operators $c_i$, $d_i^*$. The two-point function is

$$\left\langle \hat{\phi}(y) \hat{\phi}^*(y') \right\rangle = \sum_i \phi^i(y) \bar{\phi}^i(y') = (y \cdot y')^{-1}. \tag{4.4}$$

Normally, we should expect to restrict this to the cone, but the intrusion of the field $\sigma$ forces us to take into account the first-order variation off the cone as well. Since the restriction to the cone of $\partial_\alpha \phi$ is determined modulo $y_\alpha \sigma$ by the restriction of $\phi$, we have

$$\left\langle \partial_\alpha \hat{\phi}(y) \hat{\phi}^*(y') \right\rangle = -y'_\alpha (y \cdot y')^{-2}, \tag{4.5}$$

$$\left\langle \partial_\alpha \hat{\phi}(y) \partial'_\beta \hat{\phi}^*(y') \right\rangle = 2(y'_\alpha y_\beta - y_\alpha y'_\beta)(y \cdot y')^{-3} - \delta_{\alpha\beta} (y \cdot y')^{-2} + y_\alpha y'_\beta f(y \cdot y'), \tag{4.6}$$

with $f$ arbitrary. We may take $f = 0$ for convenience.

Equations (4.1), (4.5), and (4.6) give us the Minkowski two-point functions involving $\Phi$ and $\sigma$. The corresponding time-ordered products are

$$\left\langle T\Phi(x)\Phi^*(x') \right\rangle = \left[ (x - x')^2 - i0 \right]^{-1}, \tag{4.7}$$

$$\left\langle T\sigma(x)\Phi^*(x') \right\rangle = \left[ (x - x')^2 - i0 \right]^{-2}, \tag{4.8}$$

$$\left\langle T\sigma(x)\sigma^*(x') \right\rangle = 0. \tag{4.9}$$

These are the propagators that are needed to develop perturbative quantum field theory. It should be noted, however, that the meaning of Eq. (4.8) is not quite clear. The distribution (generalized function)

$$\left[ (x - x')^2 - i0 \right]^{-d/2}. \tag{4.10}$$

is analytic in $d$, except for poles at $d = 4, 6, \ldots$. Near the pole at $d = 4$ we have [19]

$$\left[ (x - x')^2 - i0 \right]^{-d/2} = \frac{2\pi^2}{d - 4} \delta(x - x') + \ldots, \tag{4.11}$$

where the unwritten terms are analytic at $d = 4$. Hence Eq. (4.8) cannot be given a precise meaning by analytic regularization.

We may attempt to resolve this by appealing to conformal invariance. Equations (4.7) and (4.8) are evidently Poincaré invariant. They are also dilatation invariant, $\Phi$ and $\sigma$ having conformal degrees 1 and 3, respectively. Under special conformal transformations the two fields transform as

$$K_\mu \Phi = \left[ x^2 \partial_\mu - 2x_\mu (x \cdot \partial + 1) \right] \Phi, \qquad (4.12)$$

$$K_\mu \sigma = \left[ x^2 \partial_\mu - 2x_\mu (x \cdot \partial + 3) \right] \sigma - \partial_\mu \Phi. \qquad (4.13)$$

Covariance of (4.7) and (4.8) is the requirement

$$\langle T\Phi(x) \partial'_\mu \Phi^*(x') \rangle = 2(x - x')_\mu \langle T\Phi(x) \sigma^*(x') \rangle.$$

This shows that $\langle T\Phi(x)\sigma^*(x') \rangle$ is fixed by covariance up to a multiple of $\delta(x - x')$ only. The singular coefficient in (4.11) can thus be replaced by a free parameter, without destroying conformal covariance. We are not sure, however, whether this is the best way to proceed. It may be that (4.11) should be taken quite literally, in the context of dimensional regularization. Another possibility that seems to offer much promise is that a careful analysis of the space of free field modes that make up the quantum field operator may lead to another method of regularizing the singular functions, in terms of unusual test function spaces. (In analytic, or dimensional regularization Schwartz spaces are used [19].) A paper by Baker and Johnson (the last paper cited in Ref. 8) seems to suggest that all these singular functions will actually cancel without any explicit appeal to regularization.

The propagators (4.7)-(4.9) are not entirely satisfactory, since (4.8) is incompatible with the free field equation $\Box \Phi = 0$. There is no simple way to cure this, but we offer two suggestions. In order to avoid repetition we present one of them here and the alternative in dealing with the case of spinor electrodynamics.

If we make use of dimensional regularization, then we may set

$$\langle \Phi(x)\Phi^*(x') \rangle = \left[ (x - x')^2 \right]^{-1-\varepsilon},$$

$$\langle \sigma(x)\Phi^*(x') \rangle = \left[ (x - x')^2 \right]^{-2-\varepsilon}, \qquad (4.14)$$

$$\langle \sigma(x)\sigma^*(x') \rangle = \frac{2}{\varepsilon} \left[ (x - x')^2 \right]^{-3-\varepsilon}.$$

This is invariant (to leading order in $\varepsilon$) under the transformations (4.12) and (4.13), and compatible with

$$\Box \Phi - 4\varepsilon \sigma = 0.$$

The required transformation properties of the six-current are

$$K_\mu J_+ = \left[ x^2 \partial_\mu - 2x_\mu (x \cdot \partial + 2) \right] J_+,$$

$$K_\mu J_\nu = \left[ x^2 \partial_\mu - 2x_\mu (x \cdot \partial + 3) \right] J_\nu + 2(x_\nu J_\mu - \delta_{\mu\nu} x \cdot J) + 2\delta_{\mu\nu} J_+, \qquad (4.15)$$

$$K_\mu J_B = \left[ x^2 \partial_\mu - 2x_\mu (x \cdot \partial + 4) \right] J_B - J_\mu.$$

A six-current that transforms correctly when (4.12) and (4.13) are applied to the fields is given by

$$J_+ = 0, \qquad J_+ = ie\Phi^*\vec{\partial}_\mu\Phi, \qquad J_B = ie(\Phi^*\sigma - \sigma^*\Phi).$$

The invariant action is

$$I[A,\Phi,\sigma] = I[A] + \int d^4x \left[ \Phi^*(\partial_\mu - ieA_\mu)^2\Phi - 2ieA_+(\Phi^*\sigma - \sigma^*\Phi) - 4e^2 A_+ A_B \Phi^*\Phi \right]$$

which is also gauge invariant under the combined transformation (3.9), and

$$\Phi \to e^{ie\Lambda}\Phi, \qquad \sigma \to e^{ie\Lambda}\left(\sigma - \tfrac{ie}{4}\Phi\Box\Lambda\right).$$

The current is conserved by virtue of the field equations. Note, however, that this includes $eA_+\Phi = 0$, obtained by variation of $\sigma^*$; this equation can be imposed only as an initial condition.

## 5. Conformal Spinor QED

Following Dirac [12] and Mack and Salam [13], we describe massless electrons in terms of an eight-spinor field $\chi$ on the six-cone, with degree of homogeneity $-2$. Dirac's wave equation is

$$\slashed{y}\slashed{\partial}\chi(y) = 0. \tag{5.1}$$

where $\slashed{y} = \Gamma_\alpha y_\alpha$, $\slashed{\partial} = \Gamma_\alpha \partial_\alpha$ and $(\Gamma_\alpha)$ is a set of six eight dimensional Dirac matrices satisfying

$$\Gamma_\alpha \Gamma_\beta + \Gamma_\beta \Gamma_\alpha = -2\delta_{\alpha\beta}.$$

The free action

$$\frac{1}{2}\int (dy)\chi^\dagger \left(\slashed{y}\vec{\slashed{\partial}} + \vec{\slashed{\partial}}\slashed{y}\right)\chi. \tag{5.2}$$

is invariant under the "Dirac transformations"

$$\chi \to \chi + \slashed{y}\sigma, \tag{5.3}$$

$\sigma$ an arbitrary spinor field of degree $-3$. The wave equation (5.1) is also invariant, as is seen by applying the identity

$$\slashed{y}\slashed{\partial} = -2\slashed{y}(3 + y\cdot\partial). \tag{5.4}$$

Physical particles are described by the space of $\chi$ fields that satisfy (5.1), modulo the subspace of solutions of the form $\chi = \slashed{y}\sigma$, the unphysical subspace.

The available choices of two-point functions and propagators for the quantum field $\hat{\chi}$ were carefully examined by Adler [14]. The general form allowed by covariance and the requirement that $\chi$ have degree of homogeneity $-2$ is

$$\langle \hat{\chi}(y)\hat{\chi}(y')\rangle = c_1 (y \cdot y')^{-2} + c_2 yy'(y \cdot y')^{-3}, \tag{5.5}$$

with arbitrary coefficients $c_1$, $c_2$. The second term is a "null propagator" that propagates only the unphysical modes, so $c_1$ must be nonzero and $c_2$ is arbitrary. Note, however, that while the second term in (5.5) satisfies the wave equation, the first term does not. This was observed by Adler [14], who nevertheless adopted

$$\langle \hat{\chi}(y)\hat{\chi}(y')\rangle = (y \cdot y')^{-2}. \tag{5.6}$$

A similar choice was made by one of us in dealing with the corresponding three-dimensional problem [20], but here it is not acceptable.

Equation (5.6) is inconsistent with the free wave equation (5.1). To ignore this problem in the present context would not be logically coherent, for we have insisted on precisely this point in the case of the free electromagnetic field. In fact, the following discussion is parallel to that of the electromagnetic case, in the last two paragraphs of Sec. 3.

As pointed out already, Eq. (5.1) admits an invariant subspace of unphysical solutions. This space is not invariantly complemented, and the physical solutions do not lie within a Gupta-Bleuler triplet. To be precise, let us abbreviate

$$D(E_0, \tfrac{1}{2}, 0) \oplus D(E_0, 0, \tfrac{1}{2}) \equiv D(E_0). \tag{5.7}$$

Then the physical solutions of (5.1) form the quotient of the nondecomposable representation

$$D(\tfrac{3}{2}) \to D(\tfrac{5}{2}), \tag{5.8}$$

the invariant subspace being unphysical. This representation cannot support a nondegenerate symplectic form. It follows that the null propagator - the second term in (5.5) - is the only propagator that is consistent with (5.1).

This situation has been met before, in the formulation of a quantum field theory of singletons on de Sitter space [21]. Our solution to the difficulty is based on that experience and especially on the work of Heidenreich [22].

What is needed is an extension of (5.8) to a Gupta-Bleuler triplet:

$$D(\tfrac{5}{2}) \to D(\tfrac{3}{2}) \to D(\tfrac{5}{2}). \tag{5.9}$$

Where can we find the extra "scalar" modes?

The answer is suggested by the fact that when differential equations seem to degenerate, and yield less regular solutions than expected, then logarithms frequently appear. Now it is easy to see, using the identity (5.4), that the two-point function

$$\langle \hat{\chi}(y)\hat{\chi}(y')\rangle = (y \cdot y')^{-2} - \tfrac{2}{\varepsilon} \slashed{y}\slashed{y}'(y \cdot y')^{-3} \ln(y \cdot y'). \tag{5.10}$$

satisfies the field equation

$$\left[\slashed{y}\slashed{\partial} - \alpha(\hat{N}+2)\right]\langle \hat{\chi}(y)\hat{\chi}(y')\rangle = 0, \qquad \hat{N} = y \cdot \partial, \qquad \alpha = 2 - \varepsilon. \tag{5.11}$$

The structure of (5.10) is clear. The physical representation $D(\tfrac{3}{2})$ is contained in the first term, but it is contaminated by the unphysical "gauge" modes, and necessarily so, since these are not invariantly complemented in (5.8).

The second term in (5.10) is made up of products of "gauge" modes and a new set of "scalar" modes. The latter are logarithmic and therefore violate the homogeneity condition. This is the price we have to pay to get a two-point function that solves the wave equation. The logarithmic modes are canonically conjugate to the unphysical modes, and their inclusion expands the field module from (5.8) to the triplet (5.9). Comparison with Ref. 21 is instructive.

We shall adopt (5.10), and pass now from Dirac's to Minkowski's notation [13]. Define ($\mu = 0,1,2,3$)

$$B = 1 + \tfrac{1}{2}(\Gamma_5 + \Gamma_4)x_\mu \Gamma_\mu, \qquad \chi(y) = x_+^{-2} B X(x),$$

where $X$ is an eight-spinor field, soon to be split into two Dirac four-spinor fields. Next

$$B^{-1}\slashed{y}\slashed{\partial} B = \tfrac{1}{2}(\Gamma_5 - \Gamma_4)\Gamma_\mu (\partial/\partial x^\mu).$$

The matrix $\Gamma_5 \Gamma_4$ has eigenvalues $+1$ and $-1$, each fourfold. We decompose the eight-spinor and the matrices $\Gamma_\mu$ accordingly:

$$X = \psi \oplus \sigma, \qquad X^\dagger = i(\psi^\dagger \oplus \sigma^\dagger)\Gamma_5, \qquad \Gamma_\mu = \gamma_\mu \oplus (-\gamma_\mu).$$

Then (5.10) gives us the propagators

$$\langle T\hat{\psi}(x)\hat{\psi}^\dagger(x')\rangle = -i\gamma \cdot (x-x')\left[(x-x')^2 + i0\right]^{-2},$$

$$\langle T\hat{\psi}(x)\hat{\sigma}^\dagger(x')\rangle = -i\left[(x-x')^2 + i0\right]^{-2}, \tag{5.12}$$

$$\langle T\hat{\sigma}(x)\hat{\sigma}^\dagger(x')\rangle = \tfrac{4}{i\varepsilon}\gamma \cdot (x-x')\left[(x-x')^2 + i0\right]^{-3}\left[\ln(x-x')^2 + x_+ + x'_+\right].$$

The appearance of $x_+$ and $x'_+$ originates in the lack of homogeneity in (5.10).

The wave equation (5.11), re-expressed in Minkowski notation, is

$$\slashed{\partial}\psi + \varepsilon \partial_+ \sigma = 0. \tag{5.13}$$

The field $\partial_+ \sigma$ is independent of $x_+$. The Lorentz condition of this theory is

$$\partial_+ \sigma = 0 \quad \text{(Lorentz condition)}. \tag{5.14}$$

It remains to be shown that, when this is imposed, as an initial condition on the physical states, it ensures the cancellation of the $x_+$ dependence of the S-matrix that is introduced by the propagators (5.12).

The transformation law for $\psi$, $\sigma$ is

$$K_\mu \psi = \left[ x^2 \partial_\mu - 2x_\mu (x \cdot \partial + 1) - \gamma_\mu x \cdot \gamma \right] \psi,$$

$$K_\mu \sigma = \left[ x^2 \partial_\mu - 2x_\mu (x \cdot \partial + 2 - \partial_+) - \gamma_\mu x \cdot \gamma \right] \sigma - \gamma_\mu \psi.$$

When (5.14) holds, the action

$$I[A, \psi, \sigma] = I[A] - \int d^4 x \left[ \tfrac{i}{2} \bar{\psi} \overleftrightarrow{\partial} \psi + e \bar{\psi} A \psi + e A_+ (\bar{\psi} \sigma + \bar{\sigma} \psi) \right]$$

is invariant, and the six-current that transform according to (4.15) is

$$J_+ = 0, \quad J_\mu = e \bar{\psi} \gamma_\mu \psi, \quad J_B = \tfrac{e}{2} (\bar{\psi} \sigma + \bar{\sigma} \psi). \tag{5.15}$$

The free wave equation, derived from the action with $e = 0$, is $\partial\!\!\!/\, \psi = 0$, whereas the equation that is satisfied by the free field operator is (5.13). The conflict is resolved when either one passes to the limit $\varepsilon = 0$ or one imposes the Lorentz condition (5.14). Varying $\psi$ and $\sigma$ with $J_B$ fixed gives

$$(\partial\!\!\!/\, - ie A\!\!\!/\,) \psi = 0. \tag{5.16}$$

and arbitrary variations of $\psi$, $\sigma$ give in addition

$$e A_+ = 0. \tag{5.17}$$

This last equation is not a field equation of the free theory, and it can be imposed only as an initial condition on the physical states. From Eqs. (3.6)-(3.8) we have

$$\Box^2 A_+ = 2 \partial \cdot J,$$

and from Eqs. (5.15) and (5.16) $\partial \cdot J = 0$, so $A_+$ is a generalized free field, as required.

Group theoretically, the extension (5.9) is interesting. While (5.8) is a representation of the double covering of SO(3,2), (5.9) is a faithful representation of the universal covering group. The requirements of field quantization (existence of a nondegenerate symplectic form) therefore leads inevitably to a logarithmic propagator. Note, however, that although we applied one method to the scalar and an alternative to the spinor, both options are available in each case.

**Appendix 1. Second Approach to the Covariant Gauge Fixing Condition in Four Dimensions**

We shall sketch here very briefly a second approach (much less conventional) to the problem of conformal covariant gauge condition in four-dimensional space-time.

We saw until now two important features coming, respectively, from the free field case and from our first approach to the interacting case:

(a) Appearance of the logarithmic term in the transformation law (2.7) for $A_\mu(x)$,

(b) Appearance of non-polynomial terms in the gauge condition (2.11) via $\rho$.

(c) A close look reveals that another alternative exists, namely, that of adding an imaginary part to the conformal degree of the electron field. The existence of such a possibility should not be too surprising: First it is known [6] that it is only for the real part of the conformal degree that we have some information from first principles - the imaginary part being completely arbitrary and generally chosen to vanish. Second, since in the case of fermions the conformal degree is half-integer, any change in sign of $\omega$ might introduce an imaginary part to the conformal degree. In this appendix we shall study in more detail a very simple conformal invariant Lagrangian that is a polynomial in the fields $A$, $\psi$, $Z$. For this Lagrangian the equations of motion admit a subsystem defined by $Z = \text{const.}$, which gives us the ordinary Maxwell-Dirac equations plus a gauge condition which is also polynomial in the fields and in their derivatives.

Let us now introduce the complex degree for the field $\psi$, a possibility mentioned briefly in Ref. 6. Suppose the transformation law under $C$ to be

$$A'^\mu(x') = \omega^2(a,x)\partial_\nu x'^\mu \left[ A^\nu(x) + \partial^\nu \ln|\omega(a,x)| Z(x) \right],$$

$$Z'(x') = Z(x),$$

$$\psi'(x') = [\omega(a,x)]^{1+ie}(1 + \not{a}\not{x})\psi(x)$$

[This transformation may be viewed as a group-consistent linkage between a space-time dependent gauge transformation of the pair $(A,Z)$ and a co-transformation of the triplet $(A,Z,\psi)$ under $C$. The gauge transformation of the pair $(A,Z)$ coincides with the usual gauge-transformation of $A$ when $Z = \text{const.}$]

Consider the Lagrangian

$$\mathscr{L} = \mathscr{L}_M + Z(x)\mathscr{L}_D - J^\mu A_\mu,$$

where

$$\mathscr{L}_M = \tfrac{1}{2}A_\mu\left(\partial_\rho F^{\rho\mu} - \tfrac{1}{2}\Box\partial^\mu Z\right) + \tfrac{1}{4}Z\Box\partial_\mu A^\mu,$$

$$\mathcal{L}_D = \frac{i}{2}\left(\tilde{\psi}\gamma^\mu \partial_\mu \psi - \partial_\mu \tilde{\psi}\gamma^\mu \psi\right).$$

Then under our co-transformation of the fields under *C* our Lagrangian $\mathcal{L}$ is a *manifestly conformal scalar* of degree 4, as can be easily checked. This last fact certainly ensures the conformal covariance of the resulting field equations:

$$\partial_\rho F^{\rho\mu} - \frac{1}{2}\Box \partial^\mu Z = J^\mu, \tag{1}$$

$$iZ\gamma^\mu \partial_\mu \psi + \frac{i}{2}\partial_\mu Z \gamma^\mu \psi + e\gamma^\mu A_\mu \psi = 0, \tag{2}$$

$$\Box \partial_\mu A^\mu + i\left(\tilde{\psi}\gamma^\mu \partial_\mu \psi - \partial_\mu \tilde{\psi}\gamma^\mu \psi\right) = 0. \tag{3}$$

The subsystem described by $Z=1$ is the Maxwell-Dirac system plus the gauge condition (3). For the "large system" (1), (2), and (3) we obtain $\partial_\mu(ZJ^\mu)=0$, so that the conserved current is $(ZJ^\mu)$ instead of $J^\mu$. This fact is not at all disturbing since the conformal degree of $Z(x)$ is anyhow zero. It is at this point that one may remark a difference between our first approach and the present one:

If we apply a gauge transformation to both $A_\mu$ and $\psi$, with gauge parameter $\phi(x)$, we obtain here for $\phi$ the equation

$$\Box^2 \phi + 2\left(\partial_\mu \phi\right)ZJ^\mu = 0, \tag{4}$$

while in the first approach we simply got the usual condition $\Box^2 \phi = 0$. In other words [though for fixed Z and $J^\mu$ satisfying the "large system" we have a linear superposition scale for the $\phi$, satisfying Eq. (A4)] the restricted gauge transformations, under which our system will be invariant, varies with Z and $J^\mu$. Such situations are not dramatic, and examples of this kind are exhibited by several supersymmetric theories.

### Appendix 2. Conformal Electrodynamics in Three Dimensions

One of the outstanding features of our space-time *being four-dimensional* is the natural conformal covariance of the (massless) equations of electrodynamics. Such a miracle does not happen in a natural way in any other dimension.

It is therefore quite interesting to know under what minimal price (if at all) one can define a conformal covariant (massless) electrodynamics in, say, three dimensions.

It is to this question that this appendix will be devoted. Before doing so, let us make the following two remarks.

(1) Several people [23] looked at three-dimensional electrodynamics from a quite different point of view: They tried to utilize "conformal symmetry breaking" in three dimensional electrodynamics (with an eventual additional term and in a certain interpretation) as a possible mechanism of mass generation in three dimensions. We shall not adopt their point of view here, although we shall include their term $*F \cdot A$ in our Lagrangian.

(2) We have another fundamental reason to be interested in electrodynamics in three dimensions: As is well known [24] there exists a very interesting phenomenon in de Sitter field theories in four dimensions. All the massless particles are composites of just two fundamental fields which play the role of preons in this theory. Each of the fundamental preon fields (or singleton fields, or Di and Rac fields as they are sometimes referred to) has a very natural interpretation as a quantum field in three dimensional space-time [21,20]. It is therefore very interesting to study fundamental interactions of quantum fields in three-dimensional space-time as a possible means of generating interacting four-dimensional quantum field theories.

Let us now come to the problem itself: Our space-time being three-dimensional [though the space is $(S_1 \times S_2)/Z_2$, we shall here concentrate on local action in Minkowski space $M_3$] the conformal group $C_3$ will be locally isomorphic to $SO_0(3,2)$. We shall consider the following Lagrangian:

$$\mathscr{L} = \mathscr{L}_M + \mathscr{L}_D + \mathscr{L}_I,$$

where

$$\mathscr{L}_M = -\tfrac{1}{4} F_{\mu\nu} F^{\mu\nu} + \frac{l}{2} *F^\mu A_\mu, \qquad F_{\mu\nu} = \partial_\mu A_\nu - \partial_\nu A_\mu, \qquad *F_\mu = \tfrac{1}{2} \epsilon_{\mu\nu\rho} F^{\nu\rho}.$$

and

$$\mathscr{L}_D = \frac{i}{2} \left( \bar\psi \gamma^\mu \partial_\mu \psi - \partial_\mu \bar\psi \gamma^\mu \psi \right), \qquad \mathscr{L}_I = -J^\mu A_\mu, \qquad \gamma^\mu \partial_\mu = \gamma^0 \partial_0 + \gamma^1 \partial_1 + \gamma^2 \partial_2.$$

and $l$ a constant. This theory has nothing of a conformal-invariant aspect. Indeed from a conformal point of view various conflicts arise between the degrees of the fields. Expected conformal degrees of $A$ and $\psi$ in three dimensions are, respectively, $\tfrac{1}{2}$ and 1. Thus $\mathscr{L}_I$, for instance, has conformal degree $\tfrac{5}{2}$ and not 3 as would have been canonical. Furthermore $\mathscr{L}_M$ does not have a well-defined conformal degree: $\mathscr{L}_M$ is the sum of two terms of different degrees. Even worse, the term $-F^2$ does not cotransform under $C_3$ as a scalar density of degree 3 (even not up to a divergence). In other words even when $l=0$ the free electromagnetic part $\mathscr{L}_M$ does not define a conformally covariant theory.

Is it possible to obtain a conformal covariant theory and if so, at what price?

Suppose that we decide to change the conformal degree of the electromagnetic potential $A$. An obvious guess would be to choose the degree $n=1$ instead of $n=\tfrac{1}{2}$. If such a choice is made one discovers that

(1) $F_{\mu\nu} F^{\mu\nu}$ is a conformal scalar of degree 4;

(2) $*F^\mu A_\mu$ is a conformal scalar of degree 3;

(3) $\mathscr{L}_I$ is a conformal scalar of degree 3.

Our guess of $n=1$ as well as points (1) and (2) are a consequence of the fact that (in any dimension of the underlying space-time) the couple $(A,F)$ transforms under $C$ in a completely reducible manner if and only if $A$ has the conformal degree 1 ($n=1$).

The only anomalous term is now $F_{\mu\nu}F^{\mu\nu}$. The way out is simple: Choose $Z$ to be an abnormal scalar field in three-dimensional space-time with conformal degree $-\frac{1}{2}$. (An abnormal *free* scalar field $\Lambda$ in $M_3$ satisfies $\Box^2\Lambda = 0$ and its degree is also $-\frac{1}{2}$.) Replace $F^2$ by $Z^2F^2$. Consider now the Lagrangian

$$\mathscr{L} = \mathscr{L}_M + \mathscr{L}_D + \mathscr{L}_I + \mathscr{L}_K$$

where $\mathscr{L}_I$ and $\mathscr{L}_D$ are as before, and where

$$\mathscr{L}_M = -\frac{Z^2}{4}F_{\mu\nu}F^{\mu\nu} + \frac{l}{2}*F^\mu A_\mu, \qquad \mathscr{L}_K = \frac{1}{4}Z\Box^2 Z.$$

Suppose that the degrees of the fields $(A,Z,\psi)$ are respectively $(1,-\frac{1}{2},1)$ with the corresponding usual cotransformation formulas under $C_3$. One then easily checks that $\mathscr{L}$ is a conformal scalar of degree 3. Consequently the field equations are covariant under $C_3$. They are

$$\partial_\mu(Z^2F^{\mu\nu}) + l*F^\nu = J^\nu,$$

$$i\gamma^\mu\partial_\mu\psi + eA_\mu\gamma^\mu\psi = 0,$$

$$\Box^2 Z - ZF^{\mu\nu}F_{\mu\nu} = 0.$$

Clearly these equations, in addition to their conformal covariance, are fully gauge invariant although $\mathscr{L}$ is only invariant up to a divergence. Clearly the current $J^\mu$ is conserved ( $\partial_\mu J^\mu = 0$) as a consequence of the equations. Its conservation implies in turn that $\in^{\nu\rho\tau}\partial_\nu F_{\rho\tau} = 0$ (for $l \neq 0$).

Let us end this appendix with the following remarks.

(a) The subsystem characterized by $Z=1$ (with eventually $l=0$) is the Maxwell-Dirac system with the additional condition $F^2 = 0$. This means that the Maxwell-Dirac subsystem will have in our system only solutions with $|\vec{E}| = |\vec{B}|$ (like plane waves). On the other hand, there will be solutions of the system with non-constant $Z$ of possibly a greater physical interest than the Maxwell-Dirac subsystem.

(b) Other approaches to conformal electrodynamics in three dimensions are possible that follow more closely the four-dimensional case treated before. We shall not elaborate on it in this thesis.

**Appendix 3. Liouville Equations**

There are some questions of direct interest to conformal invariant theories that we will treat here. For example, it is well known that in four-dimensional space-time the only conformal-invariant polynomial self-interacting scalar field theory is the massless $\lambda\phi^4$ theory. In three dimensions the corresponding theory will be the massless $\lambda\phi^6$ theory. (See the Rac field theory [21].) What happens in two dimensions? Besides the fact that in two dimensions the conformal group is infinite [a question with which we shall not deal here, our main concern being the group $C_2$ locally isomorphic to SO(2,2)], there does not exist a polynomial self-interacting scalar field theory in two dimensions since the normal conformal degree of a scalar field in two dimensions is 0. What self-interacting conformal scalar field theory can exist in two-dimensional space-time (if at all)? This brings us to the Liouville equation in two dimensions. From there we can generalize and find a new type of Liouville kind of equation of arbitrary space-time dimension; and in particular we find a *self-interacting* conformal invariant dipole-ghost in four dimensions. We remember that the free dipole ghost occurs in and is typical of some four-dimensional conformal invariant theories. The self-interacting ghost might serve as a clue for building future interacting conformal-invariant four dimensional theories (like nonlinear conformal gravity). We remark that it is sometimes useful to extend the notion of conformal invariance, by permitting in addition to the usual conformal transformation on the fields a general but fixed nonlinear transformation on the field algebra.

It is in this sense that the recently studied Liouville equation in two-dimensional space-time (in connection with completely integrable systems and with the motion of its singularities) is conformally covariant. We now come to the subject itself.

Let $S$ be a real $d$-dimensional vector space endowed with a constant metric $g$. Let $\phi$ be a real scalar field on $S$. An equation of the type

$$\Box^k \phi = \lambda e^\phi.$$

will be said to be of the Liouville type. [$k$ is here a positive integer, $\lambda$ a real constant, and $\Box = g^{\mu\nu} \left(\partial/\partial x^\mu\right)\left(\partial/\partial x^\nu\right)$.]

Some Liouville-type equations, describing a self-interacting field (especially the case $k=1$ in two dimensions) are known to be conformal invariant in a certain sense. We now prove the following proposition.

The equation $\Box^k \phi = \lambda e^\phi$ is conformally covariant if and only if $k = d/2$, provided that by covariance we mean the usual conformal covariance *composed with a fixed nonlinear transformation on the fields*.

In particular if $d = 4$, the equation $\Box^2 \phi = \lambda e^\phi$ is covariant under $C_2$.

In cases where $d$ is odd, our result does not make a direct sense, but in Euclidean space where $\Box^{d/2}$ is a pseudo-differential operator defined by a Fourier transform, the result still holds.

One proves the proposition by the nonlinear transformation $\psi = e^\phi$ which transforms our original equation into $\Box^k \ln \psi = \lambda \psi$.

When $x$ is transformed under $C_2$ suppose that

$$\psi(x) \to \psi'(x') = \omega(a,x)^n \psi(x),$$

where as usual

$$\omega(a,x) = 1 + 2(a \cdot x) + a^2 x^2$$

This means that

$$\ln \psi(x) \to \ln \psi'(x') = \ln \psi(x) + n \ln \omega(a,x)$$

[in the 0 neighborhood of $S$ where $\omega(a,x) > 0$]. Counting degrees we obtain the necessary condition for covariance: $n = 2k$.

The rest of the demonstration is simple and makes use of the following two facts.

(a) Let $d$ be even, $k$ positive integer, and $\Box'^k$ the transform of $\Box^k$ under $C_2$. Then (in $d$ dimensions) we have $\Box'^k = \omega(a,x)^{k+\frac{d}{2}} \Box^k \omega(a,x)^{k-\frac{d}{2}}$.

(b) For even $d$ and again in $d$ dimensions we have $\Box^{d/2} \ln \omega = 0$.

We end this appendix with the suggestion that it would be interesting to study both the solutions and singularities of the Liouville-type equations in $d$ dimensions (the case where $k = d/2$) in order to understand what singularities and what differences they have compared with those of the classical Liouville equation in two dimensions ($d = 2$).

**References to Section A**

# B. CONFORMAL QUANTUM YANG-MILLS

## 1. Introduction

Conformal invariance of nontrivial QED [1-5] introduces an extra scalar field $A_+$, which corresponds, in the classical theory, to a Lagrange multiplier needed to fix the gauge in a conformally invariant way [5]. Setting $A_+ = 0$ destroys the covariance of the theory or makes it equivalent to a trivial one that has only longitudinal photons. The five-component vector potential $(A_\mu, A_+)$ forms a non-decomposable representation of the conformal group. In the presence of interaction, $A_+$ couples to an extra current component of degree 4, denoted by $J_-$. The formal gauge-invariant, minimal coupling introduces unphysical field components in $J_-$. Taking $J_- = 0$ violates conformal invariance; therefore, we intend to deal with these unphysical fields and quantize them along with the physical ones.

Recently [5,6], it was shown that these unphysical components are part of a gauge field whose physical subspace is the matter field. This gauge phenomenon reinforces the importance of conformal symmetry in quantum field theory since gauge theories are the most successful ones in describing the interactions of elementary particles. Moreover, the appearance, in a natural way, of these new field components introduces new perturbative diagrams that may have a positive outcome in the renormalization program. Therefore, at this point in the development of the theory, a careful study of renormalization can be very fruitful. Another pressing point is the following: believing in the benefits of the foregoing analysis, which is brought about by requiring conformal invariance of the interaction, then looking at a gauge theory with self-coupling under the same requirements may produce interesting results even on a deeper level. As an example of such a theory, we deal in this paper with conformal Yang-Mills having local SU(N) gauge symmetry. Again, we find that unphysical components of higher degree appear as part of a current that is generated by self-interaction and couples to $A_+$. We believe that this phenomenon of field components doubling is a general property of conformal charged fields. For example, in Sec. 2, it is shown that the conformal electromagnetic potential describes the photon without the need for doubling. However, when the vector potential is charged and self-coupled, new field components surface to enlarge the gauge structure and make it richer than in the Abelian case. The appearance of these new components in the charged fields can be viewed as a recasting of the anomaly in the conformal degree of these fields into higher degree components. It also suggests an alternative to our definition of the unphysical components in which they are multiplied by a weight that depends on the coupling constant and vanishes when the coupling goes to zero. Moreover, the fact that these unphysical fields always couple to $A_+$ still holds and is very interesting when taken at the level of minimal breaking together with the remarkable property that $\langle 0|A_+(x)A_+(x')|0\rangle$ is a constant. In this context, minimal breaking means that the vacuum is not invariant under the action of special conformal transformation but still dilatation invariant (hence, no mass generation is involved). Although it is not shown here, it is believed that the new unphysical components will not contribute to the unitarity of the theory while playing a major role in the renormalization.

In this part of the thesis, we employ the manifest conformal invariance formalism [7] modified by defining an extension off the Dirac six-cone, which was introduced by us [5] and independently by Ichinose [8]. In Sec. 2, we define this extension and show that using this formalism we recover conformal QED. The covariant propagators and the underlying non-decomposable representation of the conformal group are found in Sec. 3. The Yang-Mills Lagrangian and nonlinear field equations are written in Sec. 4, where we also outline the path integral formulation and Becchi-Rouet-Stora (BRS) invariance.

The manifest formalism is based on the isomorphism of the conformal group $\mathscr{C}$ to SO(4,2). Therefore, the action of $\mathscr{C}$ in Minkowski space can be linearized by the action of SO(4,2) in $R^6$, with coordinates $\{y^a\}$, $a = 0,1,2,3,4,5$, and preserving $y^2 = y_0^2 - \vec{y}^2 + y_5^2$. The two extra dimensions are subsequently eliminated by a constraint ($y^2 = 0$) and a projection ($\lambda y \simeq y$, for $\lambda \neq 0$). The result is the projective Dirac cone - the compactified Minkowski space. Minkowski space is a dense open submanifold whose complement is the light cone at $\infty$, and with coordinates $x^\mu$ defined in the transformation

$$\left(x^\mu, x^+, x^-\right) \equiv \left(y^\mu/y^+, \ln y^+, y^2/(y^+)^2\right),$$

where $y^\pm = y^5 \pm y^4$. In the coordinates $y^\mu$, $y^\pm$, the metric

$$(\eta_{\alpha\beta}) = (\eta_{\mu\nu}) \oplus \begin{pmatrix} 0 & \tfrac{1}{2} \\ \tfrac{1}{2} & 0 \end{pmatrix},$$

where $\eta_{\mu\nu} = \mathrm{diag}(+---)$ and $\alpha, \beta = 0,1,2,3,+,-$.

If $\{L_{\alpha\beta} = -L_{\beta\alpha}\}$ is a basis for the algebra so(4,2), then the generators of the conformal group are represented by

$$\left(J_{\mu\nu}, P_\mu, K_\mu, D\right) \to \left(L_{\mu\nu}, 2L_{-\mu}, 2L_{+\mu}, 2L_{+-}\right),$$

where $(J_{\mu\nu}, P_\mu)$ are the Poincaré group generators and $D$, $K_\mu$ are, respectively, the generators of dilatation and conformal boosts.

Minimal weight, $K$-finite, irreducible representations of SO(4,2) are denoted by $D(E_0, j_1, j_2)$, where $E_0$ is the "conformal energy" and $j_1 - j_2$ is the helicity. Please see Ref. 1 for details. The conformal scalar and spinor fields carry the following nondecomposable representations (Gupta-Bleuler triplets) of $\mathscr{C}$, respectively [5]:

$$D(3,0,0) \to D(1,0,0) \to D(3,0,0), \tag{1.1}$$

$$D\left(\tfrac{5}{2}, \tfrac{1}{2}, 0\right) \to D\left(\tfrac{3}{2}, 0, \tfrac{1}{2}\right) \to D\left(\tfrac{5}{2}, \tfrac{1}{2}, 0\right). \tag{1.2}$$

and its helicity conjugate. The arrows denote semidirect sums referred to as "leaks."

## 2. Conformal QED Off the Cone

The incompatibility of conformal invariance with unrestricted gauge invariance in the five-component electrodynamics is evident in the work of many authors [1-5,9]. The scalar $A_+$ is introduced at the level of gauge fixing, but not before. We intend to deal with this problem in the "extended manifest formalism."

On the projective cone, the electromagnetic action and wave equation are [1]

$$S[a,j] = \int (dy)\left[\tfrac{1}{2} a^\alpha \partial^2 a_\alpha - a \cdot j\right], \qquad (dy) = d^6 y\, \delta(y^2), \qquad \partial^2 a_\alpha = j_\alpha. \qquad (2.1)$$

In Minkowski notations, they read

$$S[A,J] = \int d^4 x \left[\tfrac{1}{2} A_\mu \Box A^\mu + 2 A_- \Box A_+ - 4 A_- \partial \cdot A - 8 A_-^2 - A_\mu J^\mu - 2 A_+ J_- - 2 A_- J_+ \right],$$

$$\Box A_\mu + 4 \partial_\mu A_- = J_\mu, \qquad (2.2)$$

$$\Box A_+ - 2 \partial \cdot A - 8 A_- = J_+,$$

$$\Box A_- = J_-,$$

where

$$a_\alpha = \frac{\partial x^\beta}{\partial y^\alpha} A_\beta \equiv e^{-x_+} \left(V^{-1}\right)_\alpha^\beta A_\beta. \qquad (2.3)$$

and $A_\pm = \tfrac{1}{2}(A_5 \pm A_4)$.

The free action is invariant under the gauge transformation

$$(A_\mu, A_+, A_-) \to (A_\mu + \partial_\mu \Lambda, A_+, A_- - \tfrac{1}{4} \Box \Lambda). \qquad (2.4)$$

if $\Lambda$ is restricted to satisfy $\Box^2 \Lambda = 0$. Unrestricted gauge invariance of the interaction $(a \cdot j)$ gives the following current conservation:

$$\partial_\mu J^\mu + \tfrac{1}{2} \Box J_+ = 0.$$

The Lorentz condition is

$$y \cdot a = A_+ = 0,$$

and $A_+$ is a dipole ghost, $\Box^2 A_+ = 0$.

The solutions of the free wave equation $\partial^2 a_\alpha = 0$ form the following Gupta-Bleuler triplet (zero-center module [10]):

$$D\left(1,\tfrac{1}{2},\tfrac{1}{2}\right) \to \begin{bmatrix} D(2,1,0) \oplus D(2,0,1) \\ \oplus D(0,0,0) \end{bmatrix} \to D\left(1,\tfrac{1}{2},\tfrac{1}{2}\right). \tag{2.5}$$

Note the presence of the identity representation in the physical sector which has no analog in (1.1) or (1.2).

At this point, we attempt to bridge the gap between conformal invariance and unrestricted gauge invariance. We start with a full gauge-invariant, conformally invariant-free action and subsequently introduce gauge fixing that recovers (2.1). We begin by defining an extension for the vector field $a_\alpha$ off the cone in a similar manner to that of the scalar field in Ref. 5. So we modify (2.3) by retaining terms up to order $y^2$ in the expansion of the fields

$$a_\alpha(y(x)) \equiv e^{-x_+}\left(V^{-1}\right)_\alpha^\beta \left[A_\beta(x_\mu) + x^- B_\beta(x_\mu)\right]. \tag{2.6}$$

The action of special conformal transformation on these fields is

$$K_\mu A_+ = \overset{0}{\nabla}_\mu A_+,$$

$$K_\mu A_\nu = \overset{1}{\nabla}_\mu A_\nu + 2\left(x_\nu A_\mu - \eta_{\mu\nu} x \cdot A\right) + 2\eta_{\mu\nu} A_+, \tag{2.7}$$

$$K_\mu A_- = \overset{2}{\nabla}_\mu A_- - A_\mu$$

$$K_\mu B_+ = \overset{2}{\nabla}_\mu B_+ - \partial_\mu A_+,$$

$$K_\mu B_\nu = \overset{3}{\nabla}_\mu B_\nu + 2\left(x_\nu B_\mu - \eta_{\mu\nu} x \cdot B\right) + 2\eta_{\mu\nu}\left(B_+ - 2A_-\right) - \partial_\mu A_\nu, \tag{2.8}$$

$$K_\mu B_- = \overset{4}{\nabla}_\mu B_- - B_\mu - \partial_\mu A_-,$$

where

$$\overset{n}{\nabla}_\mu = x^2 \partial_\mu - 2x_\mu\left(x \cdot \partial + n\right).$$

Therefore, due to (2.6)-(2.8), the object

$$F_{\alpha\beta} = \partial_\alpha a_\beta - \partial_\beta a_\alpha$$

is a well-defined tensor on the cone - the electromagnetic tensor. Using this tensor, we can construct the following conformally invariant-free action

$$\int (dy)\left(-\tfrac{1}{4} F_{\alpha\beta} F^{\alpha\beta}\right). \tag{2.9}$$

One has to be careful when doing integration by parts, since the integration measure $(dy)$ contains $\delta(y^2)$. Therefore, it is better to write the expressions in Minkowski notation first, then do the manipulation later.

The action (2.9) is invariant under the unrestricted gauge transformation

$$a_\alpha \to a_\alpha + \partial_\alpha \omega$$

where $\omega(y(x)) = \Lambda(x_\mu) + x^- \chi(x_\mu)$. This transformation is equivalent to the following:

$$(A_\mu, A_+, A_-) \to (A_\mu + \partial_\mu \Lambda, A_+, A_- + \chi),$$

$$(B_\mu, B_+, B_-) \to (B_\mu + \partial_\mu \chi, B_+, B_-)$$

The first set is exactly that in (2.4) if we set $\chi = -\frac{1}{4}\Box\Lambda$ (i.e., $\partial^2 \omega = 0$). Gauge invariance of the interaction gives the following set of conservation laws: $J_+ = 0$, $\partial_\mu J^\mu = 0$. Now we propose to fix the gauge in (2.9) by adding the usual gauge fixing term $-\frac{1}{2}(\partial \cdot a)^2$ plus some possible invariant piece $\tilde{\mathscr{L}}$:

$$S_0[a] = \int (dy)\left[-\tfrac{1}{4}F_{\alpha\beta}F^{\alpha\beta} + \mathscr{L}_{GF}\right],$$

where

$$\mathscr{L}_{GF} = -\tfrac{1}{2}(\partial \cdot a)^2 + \tilde{\mathscr{L}}. \tag{2.10}$$

In Minkowski notation, this action takes the form

$$S_0[A,B] = \int d^4x \left[\mathscr{L}_0 + \tilde{\mathscr{L}} - 2A_+ \partial \cdot B - 2B_+(\partial \cdot A + 4A_-)\right],$$

where $\mathscr{L}_0$ is the free Lagrangian in (2.2) with $J_\pm^\mu = 0$. Using the conformal transformation (2.7) and (2.8), one can evidently show that the part of the action written explicitly is invariant. So we make the choice

$$\tilde{\mathscr{L}} = 2A_+ \partial \cdot B + 2B_+(\partial \cdot A + 4A_-), \tag{2.11}$$

in which case we recover the free action in (2.1) or (2.2). So we conclude that in the Abelian case it is possible to write an action with built-in gauge fixing, which is intrinsic on the projective cone, and no auxiliary fields ($B_\alpha$) are needed for covariant quantization. However, in the non-Abelian theory this is not the case, as we shall see in Sec. 4.

## 3. The Homogenous Propagators

The action (2.1) is invariant under the following "gauge" transformation, which we will refer to as the "cone-gauge" (c-gauge, for short):

$$a_\alpha \to a_\alpha + y^2 b_\alpha,$$

where $b_\alpha$ is an arbitrary vector field of degree 3. This is because on the cone ($y^2 = 0$)

$$\partial^2 (a_\alpha + y^2 b_\alpha) = \partial^2 a_\alpha + 4(y \cdot \partial + 3) b_\alpha = \partial^2 a_\alpha.$$

In fact, the solutions (2.5) of the free wave equation form the quotient of the non-decomposable representation of $\mathscr{C}$

$$\{\text{CQED}\} \to \{D(3,\tfrac{1}{2},\tfrac{1}{2}) \to D(4,0,0)\},$$

where {CQED} is the module (2.5) and the invariant "c-gauge" subspaces are all of the form $y^2 b_\alpha$. To find the extension to the full triplet, we investigate the free propagator $K_{\alpha\beta}(y \cdot y') = \langle a_\alpha(y) a_\beta(y') \rangle$, where the $a_\alpha(y)$ are the quantum field operators. An immediate candidate for this propagator is suggested by comparing the free wave equation with that of the scalar ($\partial^2 \Phi = 0$). That is, we set $K_{\alpha\beta} = -\eta_{\alpha\beta} K$, where $K$ is the scalar propagator [5]:

$$K = (y \cdot y')^{-1} + \lambda y^2 y'^2 (y \cdot y')^{-3},$$

and $\lambda$ is a dimensionless real parameter. Then, $K_{\alpha\beta}$ carries the representation

$$D_6 \otimes [D(3,0,0) \to D(1,0,0) \to D(3,0,0)], \tag{3.1}$$

where $D_6$ is the finite-dimensional vector representation of SO(4,2).

However, this propagator is not "clean," in the sense that it contains "spectator ghosts" carrying representations that are not Weyl-equivalent [1] to the physical ones and appear as direct sum representations.

*Proof*: Let $\hat{C}_{\alpha\beta}$ be the second order Casimir operator in the vector representation. Then $\hat{C}_{\alpha\beta} - c\eta_{\alpha\beta} = 0$ in an irreducible representation for some constant $c$. However, if the representation is non-decomposable, then $(\hat{C} - c)$ is nilpotent and only $(\hat{C} - c)^n = 0$ for some positive integer $n$ less than or at most equal to the number of levels of leak in the representation. One can easily check that

$$\left[(\hat{C} - c)^n\right]_\alpha^\beta K_{\beta\gamma} = \left[(\hat{C} - c)^n\right]_{\alpha\gamma} K \neq 0, \quad \forall \{n, c\}.$$

In our search for the free propagator, we require that it satisfies the following conditions: (i) SO(4,2) invariant; (ii) homogeneous of degree 1 and linear in $y^2$; (iii) contains the physical and Weyl-equivalent representations; and (iv) satisfies the Casimir equation $\left[(\hat{C} - c)^n\right]_\alpha^\beta K_{\beta\gamma}^{(n)} = 0$. These requirements fix $K_{\alpha\beta}$ uniquely modulo c-gauge. The third one implies that $c = 0$, since the physical representation $D(2,1,0) \oplus D(2,0,1)$ is Weyl equivalent to $D(0,0,0)$ as seen in (2.5). The following propagators are solutions to the Casimir equation:

$$K^{(1)}_{\alpha\beta} = y^2 y'^2 y_\alpha y'_\beta (y \cdot y')^{-4},$$

$$K^{(2)}_{\alpha\beta} = y^2 y'^2 \left[ \eta_{\alpha\beta} (y \cdot y')^{-3} + \left( \xi y_\alpha y'_\beta - y'_\alpha y_\beta \right)(y \cdot y')^{-4} \right], \tag{3.2}$$

$$\begin{aligned} K^{(n)}_{\alpha\beta} &= \eta_{\alpha\beta}(y \cdot y')^{-1} + \tfrac{1}{3}\left( y_\alpha y'_\beta - y'_\alpha y_\beta \right)(y \cdot y')^{-2} \\ &\quad + y^2 y'^2 \left\{ \lambda \eta_{\alpha\beta}(y \cdot y')^{-3} + \left[ \xi y_\alpha y'_\beta - \left( \lambda + \tfrac{1}{3} \right) y'_\alpha y_\beta \right](y \cdot y')^{-4} \right\}, \quad n \geq 3 \end{aligned} \tag{3.3}$$

where $\lambda$ and $\xi$ are arbitrary constants whose values can be chosen later to eliminate the most singular distributions.

It is evident that $K^{(2)}_{\alpha\beta}$ is the "c-gauge" propagator, since it is made up of modes all of the form $y^2 b_\alpha$. The nontrivial propagator (which contains a nonvanishing transverse part) is $K^{(n)}_{\alpha\beta}$ in (3.3). Therefore, the free propagator is determined, up to a "null propagator," by taking $K_{\alpha\beta} = K^{(3)}_{\alpha\beta}$.

It is interesting to note that restricting (3.3) to the cone, we recover the propagator for "gradient-type" gauge theory found by Binegar *et al.* [1] in describing conformal QED, and denoted by $K^{q-}_{\alpha\beta}$. Here, $q = 2/3$, which is just the right value needed to remove one of the two ghosts in (3.1). The propagator (3.2) without $y^2 y'^2$ factor is what they call the propagator for "current-type" gauge theory, $K^{q+}_{\alpha\beta}$ (here $q = \xi$), which carries the triplet

$$D(4,0,0) \to D(3,\tfrac{1}{2},\tfrac{1}{2}) \to D(4,0,0).$$

Now $K_{\alpha\beta}$ satisfies the following equations (mod $y^2$):

$$\partial^\alpha K_{\alpha\beta} = 0, \tag{3.4}$$

$$y^\alpha \partial^2 K_{\alpha\beta} = 0. \tag{3.5}$$

These conditions actually eliminate the "junk" in (3.1). Analysis of the modes in the Fourier expansion of $K_{\alpha\beta}$ shows that the triplet (2.5) forms the center of a larger Gupta-Bleuler triplet:

$$D(4,0,0)$$
$$\downarrow$$
$$D\left(3,\tfrac{1}{2},\tfrac{1}{2}\right)$$
$$\downarrow$$

$$\underbrace{D\left(1,\tfrac{1}{2},\tfrac{1}{2}\right) \to \begin{bmatrix} D(2,1,0) \oplus D(2,0,1) \\ \oplus D(0,0,0) \end{bmatrix} \to D\left(1,\tfrac{1}{2},\tfrac{1}{2}\right)} \qquad (3.6)$$

$$\downarrow$$
$$D\left(3,\tfrac{1}{2},\tfrac{1}{2}\right)$$
$$\downarrow$$
$$D(4,0,0)$$

where the representation $A_\alpha$ in (2.6) carries the top $\tfrac{2}{3}$ of the triplet. The free propagator (3.3) is equivalent to the following set of non-vanishing two-point functions ($\lambda + \tfrac{1}{3} = \xi = -\tfrac{1}{2}$):

$$\langle A_\mu(x) A_\nu(x') \rangle = r^{-2}\left(-\eta_{\mu\nu} + \tfrac{2}{3} r_\mu r_\nu r^{-2}\right),$$

$$\langle A_\mu(x) A_+(x') \rangle = \tfrac{2}{3} r_\mu r^{-2}, \qquad \langle A_\mu(x) A_-(x') \rangle = \tfrac{1}{3} r_\mu r^{-4},$$

$$\langle A_+(x) A_-(x') \rangle = -\tfrac{1}{3} r^{-2}, \qquad \langle A_+(x) A_+(x') \rangle = \tfrac{1}{3}. \qquad (3.7)$$

$$\langle B_\mu(x) B_\nu(x') \rangle = \tfrac{4}{3} \eta_{\mu\nu} r^{-6}, \qquad \langle B_+(x) B_-(x') \rangle = \tfrac{2}{3} r^{-6}. \qquad (3.8)$$

$$\langle A_\mu(x) B_\nu(x') \rangle = r^{-4}\left(-\eta_{\mu\nu} + \tfrac{4}{3} r_\mu r_\nu r^{-2}\right),$$

$$\langle A_\mu(x) B_+(x') \rangle = \langle B_\mu(x) A_+(x') \rangle = \tfrac{2}{3} r_\mu r^{-4},$$

$$\langle A_\mu(x) B_-(x') \rangle = \langle B_\mu(x) A_-(x') \rangle = \tfrac{2}{3} r_\mu r^{-6},$$

$$\langle A_+(x) B_-(x') \rangle = -\tfrac{1}{3} r^{-4}. \qquad (3.9)$$

where $r_\mu = x_\mu - x'_\mu$.

## 4. Conformal Yang-Mills

Conformal Yang-Mills was considered by Zaikov [11], who also found it necessary to introduce auxiliary fields. Fradkin and Palchik [12] found a nonlinear, nonlocal transformation of the fields. In this work, we deal with conformal Yang-Mills in the extended manifest formalism.

The problem here is the same as the one in conformal scalar and spinor QED. The free Lagrangian is "intrinsic" on the cone, which means that the unphysical fields do not contribute, but the interaction is not.

The general form of the pure Yang-Mills Lagrangian is

$$\mathscr{L} = \mathscr{L}_0 + g\mathscr{L}_1 + g^2\mathscr{L}_2,$$

where $g$ is a dimensionless coupling constant, $\mathscr{L}_0$ is quadratic, $\mathscr{L}_1$ cubic, and $\mathscr{L}_2$ quartic in the field. The $\mathscr{L}_0$ and $\mathscr{L}_2$ can be written intrinsically on the cone, but not $\mathscr{L}_1$. The general form of $\mathscr{L}_1$ is $(a^\alpha a^\beta D_\alpha a_\beta)$, where $D_\alpha$ is a linear differential operator of degree 1. The only such operator intrinsic on the cone is $D_\alpha = y_\alpha \partial^2 - 2(y\cdot\partial + 2)\partial_\alpha$, which is not satisfactory since it does not reproduce the three-gluon vertex, while $D_\alpha = \partial_\alpha$ does. Therefore, an extension off the cone is needed and the unphysical field $B_\alpha$ will contribute. So we adopt (2.6)-(2.8), where as usual the vector potential $a_\alpha$ is in the adjoint representation of SU(N):

$$a_\alpha(y) = a_\alpha^i(y)T_i, \qquad i = 1,\ldots,(N^2-1).$$

The $\{T_i\}$ are the Hermitian generators of SU($N$) with a Lie algebra and normalization

$$[T_i, T_j] = iC_{ij}{}^k T_k, \qquad \mathrm{Tr}(T_i T_j) = \tfrac{1}{2}\delta_{ij}.$$

Define the covariant derivative and its commutator

$$\mathscr{D}_\alpha = \mathbb{I}\partial_\alpha + iga_\alpha,$$

$$F_{\alpha\beta} \equiv (1/ig)[\mathscr{D}_\alpha, \mathscr{D}_\beta] = \partial_\alpha a_\beta - \partial_\beta a_\alpha + ig[a_\alpha, a_\beta],$$

where $\mathbb{I}$ is the unit matrix.

The pure Yang-Mills invariant action is

$$\begin{aligned} S[a] &= \int (dy)\mathrm{Tr}\left[-\tfrac{1}{4}F_{\alpha\beta}F^{\alpha\beta}\right] \\ &= \int (dy)\mathrm{Tr}\left\{-\tfrac{1}{2}\partial_\alpha a_\beta\left(\partial^\alpha a^\beta - \partial^\beta a^\alpha\right) - ig\left[a^\alpha, a^\beta\right]\partial_\alpha a_\beta + \tfrac{1}{4}g^2\left[a_\alpha, a_\beta\right]\left[a^\alpha, a^\beta\right]\right\} \end{aligned} \qquad (4.1)$$

$$S[A,B] = \int d^4x\left[(\mathscr{L}_0 - \mathscr{L}_{GF}) + g\mathscr{L}_1 + g^2\mathscr{L}_2\right], \qquad (4.2)$$

where $\mathscr{L}_{GF}$ is the trace of (2.10) and (2.11), and in the free theory $\mathscr{L}_{GF} = \tilde{\mathscr{L}}$ since $\partial\cdot a = 0$ as shown in (3.4):

$$\mathcal{L}_0 = \mathrm{Tr}\left(\tfrac{1}{2}a^\alpha \partial^2 a_\alpha\right) = \mathrm{Tr}\left(\tfrac{1}{2}A^\mu \Box A_\mu + 2A_- \Box A_+ - 4A_- \partial \cdot A - 8A_-^2\right)$$

$$\mathcal{L}_1 = -i\mathrm{Tr}\{[A^\mu, A^\nu]\partial_\mu A_\nu + 2[A^\mu, A_+]\partial_\mu A_- + 2[A^\mu, A_-]\partial_\mu A_+$$
$$+ 2A_+[A^\mu, B_\mu] + 4A_+[A_-, B_+]\}$$

$$\mathcal{L}_2 = Tr\{\tfrac{1}{4}[A^\mu, A^\nu][A_\mu, A_\nu] - 2[A_+, A_-]^2 + 2[A^\mu, A_+][A_\mu, A_-]\}$$

Each term in (4.2) is separately invariant. Note, however, that the unphysical fields $B_\alpha$ that appear in $\mathcal{L}_1$ cannot be extracted into an invariant piece as in the Abelian case.

The action (4.1) is invariant under the gauge transformation

$$a_\alpha \to \Omega a_\alpha \Omega^{-1} - \tfrac{i}{g}\Omega\left(\partial_\alpha \Omega^{-1}\right), \qquad (4.3)$$

where

$$\Omega = e^{-ig\omega}, \quad \omega(y(x)) = \left[\Lambda^i(x_\mu) + x^- \chi^i(x_\mu)\right]T_i.$$

The infinitesimal form of (4.3) is

$$\delta a_\alpha = \partial_\alpha \omega + ig[a_\alpha, \omega] = [\mathcal{D}_\alpha, \omega].$$

In the Abelian case, there is no self-coupling and the free field, by virtue of Eq. (3.4), satisfies $\partial \cdot a = 0$. Therefore, in the gauge transformation $\delta a_\alpha = \partial_\alpha \omega$, this requires $\partial^2 \omega = 0$, which reads $\chi = -\tfrac{1}{4}\Box\Lambda$; in agreement with the choice made in Sec. 2 for QED.

The infinitesimal gauge transformation amounts to the following:

$$\delta A_\mu = \partial_\mu \Lambda + ig[A_\mu, \Lambda] = [D_\mu, \Lambda], \quad \delta A_+ = ig[A_+, \Lambda], \quad \delta A_- = \chi + ig[A_-, \Lambda],$$

$$\delta B_\mu = ig[B_\mu, \Lambda] + [D_\mu, \chi], \qquad \delta B_\pm = ig\{[B_\pm, \Lambda] + [A_\pm, \chi]\},$$

and $D_\mu = \mathbb{I}\partial_\mu + igA_\mu$.

We fix the gauge in (4.1) and (4.2) by adding $\mathcal{L}_{\mathrm{GF}}$ and get the following conformally covariant nonlinear equations upon variations of $A_\alpha$:

$$[D^\nu, f_{\nu\mu}] + \partial_\mu \partial \cdot A + 4\partial_\mu A_- - 2ig\{[A_+, [D_\mu, A_-]] + [A_-, [D_\mu, A_+]] + [B_\mu, A_+]\} = J_\mu, \qquad (4.4a)$$

$$[D^\mu, [D_\mu, A_-]] - ig\{[A_\mu, B^\mu] + 2[A_-, B_+]\} + 2g^2[A_-, [A_-, A_+]] = J_-, \qquad (4.4b)$$

$$[D^\mu, [D_\mu, A_+]] - 2\partial \cdot A - 8A_- + 2ig[A_+, B_+] + 2g^2[A_-, [A_+, A_-]] = J_+, \qquad (4.4c)$$

where $f_{\mu\nu} = \partial_\mu A_\nu - \partial_\nu A_\mu + ig[A_\mu, A_\nu]$, and sources have been included. Variations with respect to $B_\alpha$ give the equation $gA_+ = 0$, which is not an equation for the free theory and can only be imposed as initial conditions on the physical subspace.

The propagators for the free quantum fields are just those given in (3.7)-(3.9) with superscripts $i, j$, and $\delta^{ij}$ multiplying the right side. The last set (3.9) is incompatible with two of the free wave equations obtained from (4.4a) and (4.4b) with $g = 0$. The reason is as was stated below (3.6) in that the $A_\alpha$ carry the "physical" modes that do satisfy the free wave equations, but they also carry the "scalar" modes that do not. To cure this, one may follow one of two procedures presented in Ref. 5. The first makes use of dimensional regularization and the second introduces logarithmic modes in the "scalar" sector. However, we will not pursue this any further in this part of the thesis since it does not pose any problem to the physics and it only amounts to a better choice of field variables that splits $A_\alpha$ into its two components, the "physical" and "scalar."

In the presence of matter, gauge invariance of the interaction $\text{Tr}(a \cdot j)$ gives the following set of conservation laws for the current:

$$J_+ = 0, \quad \text{and} \quad [D_\mu, J^\mu] + 2ig[A_+, J_-] = 0. \tag{4.5}$$

The spinor action and six-current are [5]

$$S[\psi, \sigma] = \int d^4x \left[ -\tfrac{i}{2} \bar\psi \slashed{\partial} \psi + g\bar\psi \slashed{A} \psi + g(\bar\psi A_+ \sigma + \bar\sigma A_+ \psi) \right],$$

$$J_+^i = 0, \quad J_\mu^i = -g\bar\psi \gamma_\mu T^i \psi, \quad J_-^i = -\tfrac{g}{2}(\bar\psi T^i \sigma + \bar\sigma T^i \psi),$$

and (4.5) is satisfied by virtue of the field equations. The action of special conformal transformation on these spinors is

$$K_\mu \psi = \overset{3/2}{\nabla}_\mu \psi - \tfrac{1}{2}[\gamma_\mu, x \cdot \gamma]\psi,$$

$$K_\mu \sigma = \overset{5/2}{\nabla}_\mu \sigma - \tfrac{1}{2}[\gamma_\mu, x \cdot \gamma]\sigma - \gamma_\mu \psi.$$

The homogeneous two-point functions are

$$\langle \psi(x)\bar\psi(x') \rangle = i(r \cdot \gamma)r^{-2}, \quad \langle \psi(x)\bar\sigma(x') \rangle = ir^{-4}, \quad \langle \sigma(x)\bar\sigma(x') \rangle = 0.$$

The total Yang-Mills Lagrangian is

$$\mathscr{L}_{\text{YM}} = \mathscr{L}_0 + g\mathscr{L}_1 + g^2\mathscr{L}_2 - \tfrac{i}{2}\bar\psi \slashed{D} \psi + g(\bar\psi A_+ \sigma + \bar\sigma A_+ \psi)$$

The new vertices are

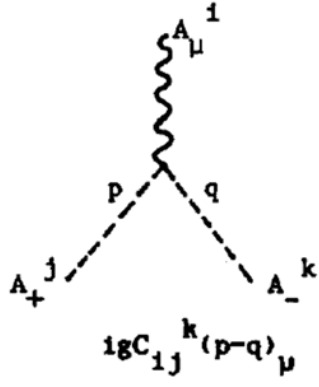
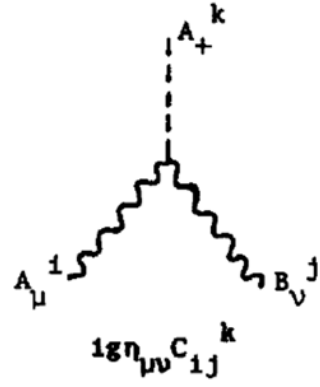
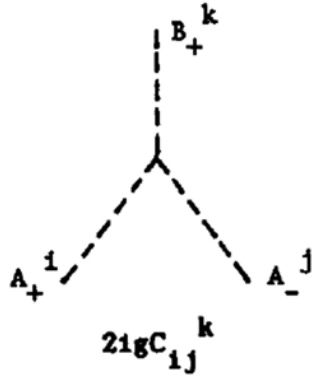
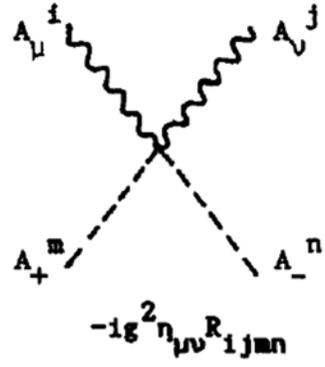
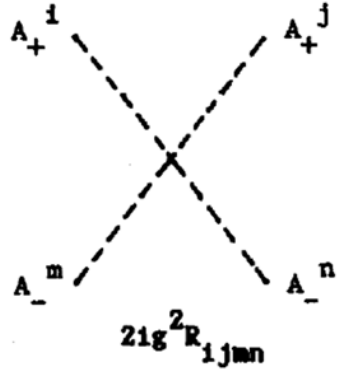
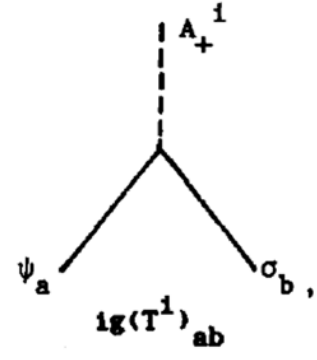

where $R_{ijmn} = C_{im}{}^{k}C_{jnk} + C_{in}{}^{k}C_{jmk}$ and $a,b = 1,...,N$.

Notice that the unphysical fields $B_\alpha$ and $\sigma$ always couple to $A_+$. It remains to be shown that the coupling of the unphysical fields do not contribute to the physical processes.

Finally, we outline the path integral formulation in one gauge using the Faddeev-Popov ansatz and indicate the residual BRS invariance.

The generating functional for the Green's functions is

$$Z[j] = \mathcal{N} \int (\mathcal{D} a_\alpha) \exp i \left\{ \int (dy) \mathrm{Tr}\left[ -\tfrac{1}{4} F_{\alpha\beta} F^{\alpha\beta} - a \cdot j \right] \right\};$$

then insert $1 = \int (\mathscr{D}\omega)\Delta_{FP}[a_\alpha^\omega]\delta(\partial \cdot a^\omega - \phi)$, which defines the Faddeev-Popov determinant $\Delta_{FP}$.

Gauge invariance leaves a meaningless infinite volume integral in gauge space that is absorbed in the normalization factor $\mathscr{N}$. Then, the constant

$$\int (\mathscr{D}\phi)\exp\left[-\tfrac{i}{2}\int (dy)\mathrm{Tr}\phi^2\right],$$

is inserted and integration over $\phi$ is carried out to result in the following amplitude:

$$Z[j] = \mathscr{N}'\int (\mathscr{D}a_\alpha)\Delta_{FP}[a]\exp i\left\{\int (dy)\mathrm{Tr}\left[-\tfrac{1}{4}F_{\alpha\beta}F^{\alpha\beta} - \tfrac{1}{2}(\partial \cdot a)^2 - a \cdot j\right]\right\}.$$

Hence $\Delta_{FP}[a] = \det\{M[a]\}$, where $M[a]$ is the operator

$$\delta(y^2)M(y,y') = \delta^6(y-y')\left(\partial^\alpha \mathscr{D}_\alpha\right).$$

Then, we express the determinant as the integral

$$\Delta_{FP}[a] = \int (\mathscr{D}\eta_i)(\mathscr{D}\eta_i^\dagger)\exp i\left\{\int (dy)\mathrm{Tr}\left(\eta^\dagger \partial^\alpha [\mathscr{D}_\alpha,\eta]\right)\right\},$$

over $2(N^2-1)$ Grassmann scalars $\eta_i$ and $\eta_i^\dagger$, referred to as FP ghosts. Finally, we obtain the generating functional

$$Z[j,k,k^\dagger] = \mathscr{N}\int (\mathscr{D}a)(\mathscr{D}\eta)(\mathscr{D}\eta^\dagger)\exp iS[a,\eta,\eta^\dagger,j,k,k^\dagger],$$

$$S = \int (dy)\mathrm{Tr}\left\{-\tfrac{1}{4}F_{\alpha\beta}F^{\alpha\beta} - \tfrac{1}{2}(\partial \cdot a)^2 - a \cdot j + \eta^\dagger \partial^\alpha [\mathscr{D}_\alpha,\eta] - \eta^\dagger K - \eta K^\dagger\right\}. \tag{4.6}$$

where we have included sources for the ghosts.

The effective action (4.6) is not invariant under the gauge transformation (4.3), but it is invariant under the rigid BRS transformations

$$\delta a_\alpha = \theta[\mathscr{D}_\alpha,\eta], \qquad \delta\eta = \tfrac{g}{2i}\theta[\eta,\eta], \qquad \delta\eta^\dagger = \theta(\partial \cdot a),$$

and $\theta$ is a Grassmann constant parameter. This invariance is used to derive the Ward identities, which may differ from Slavnov-Taylor identities for ordinary Yang-Mills. But this question is beyond the scope of this thesis. We also leave, for a possible future study, the question of choosing a better gauge fixing term (e.g., $\mathscr{L}_{GF}$) and the resulting BRS invariance. Therefore, no attempt is made here to write (4.6) in Minkowski notation and further the investigation.

# C. THE SPINOR GAUGE FIELD AND CONFORMAL QED

## 1. Introduction

The gauge is a concept that can be defined in two settings, by differential geometry or by group theory. The latter is wider and more general since all known gauge theories (conventional and unconventional [1]) fall in its domain. While in differential geometry the emphasis is placed on general covariance under the group of diffeomorphisms, in group theory the gauge concept relies on the existence of nontrivial cohomology of extensions of representations.

Gauge theories are universally characterized by non-decomposable representations of the underlying space-time symmetry group. These occur at the limit of unitarity where the representation becomes singular and falls into a chain of sub-representations each of which is not invariantly complemented. Therefore, the problem is to construct a non-decomposable representation from irreducible ones. As an example, let $M_1$ and $M_2$ be two representations of a Lie group $G$ acting in the topological vector spaces $V_1$ and $V_2$ respectively, and let $N \in \mathscr{L}$ be a linear map from $V_2$ to $V_1$. We want to build a representation on $V = V_1 \oplus V_2$ which is not equivalent to $M_1 \oplus M_2$ and in which $g \in G$ acts as

$$\begin{pmatrix} M_1(g) & N(g) \\ 0 & M_2(g) \end{pmatrix} \begin{pmatrix} x_1 \\ x_2 \end{pmatrix}, \qquad x_i \in V_i.$$

The condition that this is a representation translates into the natural action of $G$ on the space of linear maps $\mathscr{L}$ from $V_2$ to $V_1$, then the inequivalent extensions are classified by the first cohomology $H^1(G, \mathscr{L})$ [2]. A necessary condition for nontriviality of this cohomology is the equality of the center of the enveloping algebra (infinitesimal character) for $M_1$ and $M_2$. The extension sequence is written as:

$$M_1 \to M_2,$$

where the arrow is a semidirect sum indicating a leak between the two parts under the action of the group and pointing towards the invariant sub-representation.

In gauge theories $V$ is a Hilbert space in which $M_2$ is irreducible and the extension has a nondegenerate indefinite Hermitian form. Therefore, it follows that $M_1$ is equivalent to a nontrivial extension $M_2 \to M_3$ [3] and the original representation becomes a symmetric triplet ($M_2 \to M_3 \to M_2$) generically called the "Gupta-Bleuler triplet" and denoted by:

$$D(s) \to D(p) \to D(g).$$

The total space $V$ has an invariant subspace $V_+$ which in turn contains an invariant subspace $V_g$. Neither of these subspaces is invariantly complemented in a true gauge theory. $V_g$ is the null radical (the "gauge" modes) of $V_+$ and carries the representation $D(g)$. The "physical" representation $D(p)$ lives in the quotient space $V_+/V_g$ on which the induced metric is definite. $V/V_+$ is the space of "scalar" modes $D(s)$ and the action of $G$ on this space is equivalent to that on $V_g$. Imposing the "Lorentz condition" will suppress these scalar modes. The words in quotations above are generic terminology for gauge theories borrowed from the classic example of indefinite-metric quantization of QED by Gupta and Bleuler in 1950 [4] which will be discussed briefly here. But, first we want to give a short account of gauge invariance in Lagrangian field theory to explain the concept of gauge fixing and how to build the representation. Gauge invariance implies that the wave operator obtained by stationary variations of the action has a large null space; that is, certain modes of the field (the gauge modes) solve the wave equation trivially. This means that the wave operator is degenerate (its kernel is "too large"). In quantum field theory, we need to define a free propagator for the quantum field. This propagator is a distribution that satisfies the inhomogeneous wave equation, which means that it is an inverse of the wave operator (defined with the help of appropriate boundary conditions). Now if the kernel of this operator is "too large" then we may not succeed to find those boundary conditions necessary for physical modes propagation. Therefore, we need to "fix" the gauge or in a better terminology, restrict the gauge. Covariant gauge fixing is usually accomplished by adding to the action a gauge dependent $G$-invariant term that vanishes on the physical subspace - i.e., the Lorentz condition term. In this way we get a new invertible wave operator whose positive energy solutions together with the requirement of one-particle interpretation (finite norm states) lead to the construction of the Hilbert space that supports the "Gupta-Bleuler triplet".

Now let us consider QED as an illustration for the concepts discussed above. The action of electromagnetism and the resulting Maxwell's equations are, respectively:

$$-\tfrac{1}{4}\int d^4 x\, F_{\mu\nu} F^{\mu\nu}\,,\quad \Box A_\mu - \partial_\mu \partial\cdot A = 0\,,$$

where $F_{\mu\nu} = \partial_\mu A_\nu - \partial_\nu A_\mu$ and the Minkowski metric $\eta_{\mu\nu} = \text{diag}(+---)$. The modes $A_\mu = \partial_\mu \Lambda$ are trivial solutions of the wave equation and the action is invariant under the gauge transformation $A_\mu \to A_\mu + \partial_\mu \Lambda$, where $\Lambda$ is an arbitrary real scalar of degree zero. Therefore, the Maxwell operator $\eta_{\mu\nu}\Box - \partial_\mu \partial_\nu$ is degenerate and the gauge has to be fixed. We choose to do that in a covariant way by adding the term $-\tfrac{1}{2}\lambda(\partial\cdot A)^2$ to the Lagrangian ($\lambda \neq 0$). For simplicity we take $\lambda = 1$ corresponding to what is known as the Feynman gauge. The resulting equation and norm are

$$\Box A_\mu = 0\,, \qquad \|A\| = -\int d^3 x \left(A^\mu i\overleftrightarrow{\partial}_0 A_\mu\right) < \infty\,. \tag{1.1}$$

The metric is indefinite as evidenced by the Minkowski metric $\eta_{\mu\nu}$. The positive energy solution spaces are:

$$V = \{A_\mu : \Box A_\mu = 0,\ \|A\| \in R\}\,,$$

$$V_+ = \{A_\mu : \Box A_\mu = 0,\ \partial\cdot A = 0,\ \|A\| \geq 0\}\,,$$

$$V_g = \{A_\mu : \Box A_\mu = 0,\ \partial \cdot A = 0,\ A_\mu = \partial_\mu \Lambda,\ \|A\| = 0\}.$$

and

$V_g$ the space of longitudinal photons

$V_+ / V_g$ the space of transverse photons

$V / V_+$ the space of scalar photons

The space of solutions of (1.1) carries a representation which is a symmetric triplet and indecomposable

$$D(0,0) \to [D(0,+1) \oplus D(0,-1)] \to D(0,0),$$

where $D(0,\lambda)$ is the massless unitary irreducible representation of the Poincaré group with helicity $\lambda$. The triplet is equivalent to the module

$$[\partial \cdot A,\ \Box \partial \cdot A = 0] \to [F_{\mu\nu},\ \partial^\mu F_{\mu\nu} = 0] \to [\Lambda,\ \Box \Lambda = 0].$$

The Lorentz condition ($\partial \cdot A = 0$) is not satisfied by the quantum field operator, it can only be imposed as an initial condition on the physical subspace ($\langle \text{Phys}|\partial \cdot A|\text{Phys}\rangle = 0$) which will be maintained when the electromagnetic potential is coupled to a conserved current $J_\mu$. That is $\Box A_\mu = J_\mu$ and current conservation $\partial \cdot J = 0$ give $\Box \partial \cdot A = 0$, so the scalar field $\partial \cdot A$ is a free field and the S-matrix will be free of scalar photons maintaining unitarity. Moreover, due to current conservation the gauge modes decouple since

$$\int d^4x\, J^\mu \partial_\mu \Lambda = -\int d^4x\, \Lambda \partial \cdot J = 0.$$

The structure discussed above is a signature of a gauge theory. The representation carried by the gauge potential occurs at the limit of unitarity and masslessness. Therefore, gauge theories are extensions (nontrivial, but not necessarily unitary) of massless field theories. Hence, one is tempted to look for possible extensions of the massless spinor field that can support a gauge structure. An indication that such an investigation might be fruitful is found in supersymmetry if, say, the electron field belongs to the gauge supermultiplet. Another input to this investigation is the observation that pure gauge theories before gauge fixing (*e.g.*, electrodynamics, Yang-Mills, *etc.*) have a space-time symmetry group which is larger than the Poincaré group, that is the conformal group *C*. Therefore, the problem translates into building a spinor representation of the conformal group which is a non-trivial extension of the massless one and carries the gauge structure.

It is known [5-7] that the conformal spinor field is, in fact, a gauge field whose physical subspace is the massless electron. The representation of *C* carried by the spinor is indecomposable and forms a symmetric triplet (the Gupta-Bleuler triplet) as shown in references [6] and [7]. A conformally covariant quantization was carried out in reference [6]; however, some difficulties associated with the quantization of the charged conformal fields arises [6,8,9]. In Section 2.1 we enumerate these problems and investigate them, then in Section 2.2 a new attempt is presented which resolves some of these problems and presents some new features. However, other difficulties emerge which call for a deeper analysis of the problem. This is carried out in Section 3, where it is realized that this situation has been encountered before in the formulation of the singleton's field theory at spatial infinity of de Sitter space [10]. Final remarks are presented for future developments in a super-conformal setting.

## 2. The Conformal Spinor

### 2.1 *The Old Model*

The model [5,6] is written in terms of an 8-component spinor $\chi$; a multiplet of two Dirac spinors $\psi$ and $\sigma$ with conformal degree $-3/2$ and $-5/2$, respectively. The action of special conformal transformation on the multiplet is indecomposable:

$$i\left[K_\mu,\psi\right] = \overset{3/2}{\nabla}_\mu \psi - \tfrac{1}{2}\left[\gamma_\mu, \slashed{x}\right]\psi,$$

$$i\left[K_\mu,\sigma\right] = \overset{5/2}{\nabla}_\mu \sigma - \tfrac{1}{2}\left[\gamma_\mu, \slashed{x}\right]\sigma - \gamma_\mu \psi.$$

where $\overset{n}{\nabla}_\mu = x^2 \partial_\mu - 2x_\mu(x \cdot \partial + n)$ and $n$ is the conformal degree; $\slashed{x} = x^\mu \gamma_\mu$. The general free action is ($\lambda$ is a real parameter)

$$\int d^4 x \left[\tfrac{i}{2}\bar{\psi}\overset{\leftrightarrow}{\slashed{\partial}}\psi + 2i\lambda(\bar{\sigma}\psi - \bar{\psi}\sigma)\right], \tag{2.1}$$

which is invariant under the following "gauge transformations":

(i)  $\lambda = 0$: $(\psi,\sigma) \to (\psi, \sigma + \Omega)$, $\Omega$ a spinor of degree $-5/2$;

(ii) $\lambda \neq 0$: $(\psi,\sigma) \to (\psi, \sigma + \zeta\psi)$, $\zeta$ a real scalar of degree $-1$.

The most general covariant and homogeneous two-point functions are:

$$\langle \psi(x)\bar{\psi}(x')\rangle = i\slashed{r}r^{-4}, \quad \langle \psi(x)\bar{\sigma}(x')\rangle = ir^{-4}, \quad \langle \sigma(x)\bar{\sigma}(x')\rangle = 2i\lambda'\slashed{r}r^{-6}, \tag{2.2}$$

where $r = x - x'$, and $\lambda'$ is a dimensionless real parameter. However, these 2-point functions are incompatible with the free wave equation obtained from (2.1) for all $\lambda$ and $\lambda'$. Adler [8] was the first to point out and wrestle with this difficulty. Other attempts to resolve this problem were presented in reference [6] either by generalizing the concept of conformal degree by introducing logarithmic modes that violates the homogeneity condition, or by making use of dimensional regularization which leaves the theory invariant only up to leading order in the infinitesimal regularizer of 4 dimensions. Another problem with (2.1) is that if $\lambda \neq 0$, then one obtains only trivial solutions to the wave equations, while if $\lambda = 0$ the action becomes degenerate as shown by its invariance under the unrestricted gauge transformation (i) above. Finally, when the spinor is coupled to the electromagnetic potential in a conformally invariant way [6], a new problem related to current conservation surfaces. The interaction Lagrangian is

$$\mathcal{L}_I = -eA^\mu(\bar{\psi}\gamma_\mu\psi) - eA_+(\bar{\psi}\sigma + \bar{\sigma}\psi)$$

and the current is conserved (for $\lambda = 0$) only on the physical subspace ($\langle\text{Phys}|A_+|\text{Phys}\rangle = 0$):

$$\partial_\mu J^\mu = ieA_+(\bar{\sigma}\psi - \bar{\psi}\sigma)$$

Another flaw in the current is that $J_0$ is not a generator of charge for the spinor $\sigma$ which is assumed to be a field with a well-defined charge as seen from the reality of the total action.

These problems can be understood when one realizes the following two points:

1. The spinor $\psi$ carries two modes: a "physical" mode that satisfies the massless Dirac equation ($\slashed{\partial}\psi = 0$) and contributes to the 2-point function $\langle\psi(x)\bar{\psi}(x')\rangle$; and a "scalar" mode that does not and contributes to $\langle\psi(x)\bar{\sigma}(x')\rangle$.

2. $J_0$ must contain $\sigma$, which implies that we must include derivatives of $\sigma$ in the free action; that is terms of the form $\bar{\rho}\slashed{\partial}\sigma$ where $\rho$ is a spinor of degree $-1/2$. Therefore, we propose a new model which is a nontrivial extension of the old one and is written in terms of the spinor triplet $\chi \equiv (\rho,\psi,\sigma)$. The implication is that $\rho$ carries the "scalar" modes ($\slashed{\partial}\psi$ above) conjugate to the "gauge" modes in $\sigma$.

### 2.2 A New Model

We propose the following general linear transformation rules for the three spinors under the action of the conformal boost:

$$i[K_\mu,\rho] = \overset{1/2}{\nabla}_\mu \rho - \tfrac{1}{2}[\gamma_\mu,\slashed{x}]\rho$$

$$i[K_\mu,\psi] = \overset{3/2}{\nabla}_\mu \psi - \tfrac{1}{2}[\gamma_\mu,\slashed{x}]\psi - \alpha\gamma_\mu\rho, \qquad (2.3)$$

$$i[K_\mu,\sigma] = \overset{5/2}{\nabla}_\mu \sigma - \tfrac{1}{2}[\gamma_\mu,\slashed{x}]\sigma - \beta\gamma_\mu\psi - \xi\partial_\mu\rho.$$

$\alpha$, $\beta$, and $\xi$ are dimensionless real parameters. The requirement that this is a representation of the conformal algebra gives $\xi = \alpha\beta$. It is a nontrivial extension of the previous case only if $\alpha \neq 0$ and $\beta \neq 0$, therefore we take $\alpha = \beta = 1$. Then the most general free Lagrangian density is the following 2-parameter invariant:

$$\mathcal{L}_\chi(\zeta,\zeta') = \tfrac{i}{2}\bar{\psi}\slashed{\partial}\psi + 2i\zeta(\bar{\sigma}\psi - \bar{\psi}\sigma) + \tfrac{i}{2}\zeta(\bar{\rho}\slashed{\partial}\sigma + \bar{\sigma}\slashed{\partial}\rho)$$
$$+ \tfrac{i}{2}(\zeta-1)(\bar{\rho}\Box\psi - \bar{\psi}\Box\rho) - \tfrac{i}{8}\zeta'\bar{\rho}\Box\slashed{\partial}\rho \qquad (2.4)$$

It is singular if $2\zeta + \zeta' = 1$. Therefore, we take $2\zeta + \zeta' \neq 1$ and obtain the following two sets of free wave equations:

–41–

(i) $\zeta \neq 0$: $\psi + \frac{1}{2}\slashed{\partial}\rho = 0$, $\sigma + \frac{1}{4}\Box\rho = 0$, $\Box\slashed{\partial}\rho = 0$, (2.5)

and the action is invariant under the gauge transformation

$$(\rho, \psi, \sigma) \rightarrow \left(\rho + \omega,\ \psi - \tfrac{1}{2}\slashed{\partial}\omega,\ \sigma - \tfrac{1}{4}\Box\omega\right),$$

where $\omega$ is a spinor parameter of degree $-1/2$ and restricted to satisfy $\Box\slashed{\partial}\omega = 0$.

(ii) $\zeta = 0$: $\slashed{\partial}\psi + \frac{1}{2}\Box\rho = 0$, $\Box\slashed{\partial}\rho = 0$, (2.6)

and the action is invariant under the following gauge transformation:

$$(\rho, \psi, \sigma) \rightarrow \left(\rho + \omega,\ \psi - \tfrac{1}{2}\slashed{\partial}\omega,\ \sigma + \Omega - \tfrac{1}{4}\Box\omega\right)$$

where $\Omega$ is a spinor of degree $-5/2$.

The first set of equations imply that all the field modes are contained in $\rho$ and it corresponds to the trivial triplet. This can be seen as follows: Starting with (2.1) of the previous model and making the field redefinition

$$\psi \rightarrow \psi + \tfrac{1}{2}\slashed{\partial}\rho,\quad \sigma \rightarrow \sigma + \tfrac{1}{4}\Box\rho$$

which transform in the right way, we obtain the action of (2.4) with $\zeta = \lambda$. However, it was shown in the above subsection that $\lambda \neq 0$ corresponds to the trivial case. Another argument against (2.5) comes from the Lorentz condition associated with the spinor gauge field. The Lorentz condition is an invariant condition that is more restrictive than the wave equations and annihilates the "scalar" mode. For (2.5) the only such possibility is $\rho = 0$ which is, obviously, unacceptable since it eliminates all the field modes. Therefore, we reject this case and pass to the second one ($\zeta = 0$). In this case, $\rho = 0$ is consistent with the wave equations. In addition, $\rho = 0$ gives $\slashed{\partial}\psi = 0$ which was shown at the end of the previous subsection to be the condition that kills the "scalar" modes. Therefore, we choose $\zeta = 0$ and obtain the following free action with $\zeta' \neq 1$:

$$\int d^4x \left[\tfrac{i}{2}\bar{\psi}\slashed{\partial}\psi + \tfrac{i}{2}\bar{\psi}\Box\rho - \tfrac{i}{2}\bar{\rho}\Box\psi - \tfrac{i}{8}\zeta'\bar{\rho}\Box\slashed{\partial}\rho\right]$$

The non-vanishing covariant 2-point functions are:

$$\langle \rho(x)\bar{\rho}(x')\rangle = i\slashed{r}r^{-2},\quad \langle \rho(x)\bar{\psi}(x')\rangle = ir^{-2},$$

$$\langle \rho(x)\bar{\sigma}(x')\rangle = i\slashed{r}r^{-4},\quad \langle \psi(x)\bar{\psi}(x')\rangle = i\slashed{r}r^{-4}.$$

It can be easily verified that the problem of incompatibility between the 2-point functions and the free wave equations which was present in the previous model is now resolved.

However, the free field module contains "spectator ghosts" that do not mix kinematically with the physical modes. They carry non-unitary representations of the conformal group that are not Weyl-equivalent to the physical ones and appear as a direct sum. This statement is proved in Appendix I using the properties of the second order Casimir operator in a non-decomposable representation. New problems also emerge when interaction is introduced. In the notation of reference [9]:

$$\mathcal{L}_I = -A_\mu J^\mu - A_+ J_- - A_- J_+,$$

where the six-current $J_\pm^\mu$ is quadratic in the spinors $(\rho,\psi,\sigma)$ and transforms under the action of special conformal transformations as:

$$i[K_\mu, J_+] = \overset{2}{\nabla}_\mu J_+,$$

$$i[K_\mu, J_\nu] = \overset{3}{\nabla}_\mu J_\nu + 2(x_\nu J_\mu - \eta_{\mu\nu} x \cdot J) + 2\eta_{\mu\nu} J_+, \quad (2.7)$$

$$i[K_\mu, J_-] = \overset{4}{\nabla}_\mu J_- - J_\mu,$$

and $\overset{n}{\nabla}_\mu$ is defined as before. This construction is carried out in Appendix II where we end up with a general nine-parameter model for $\mathcal{L}_I$. Then, together with the usual gauge transformation of the vector potential,

$$\delta(A_+, A_\mu, A_-) = (0, \partial_\mu \Lambda, -\tfrac{1}{4}\Box\Lambda).$$

We associate the following general and covariantly extended gauge transformation for the spinors:

$$\delta\rho = ie\Lambda\rho, \qquad \delta\psi = ie\Lambda\psi + ie\alpha\,\slashed{\partial}\Lambda\rho,$$

$$\delta\sigma = ie\Lambda\sigma + ie\beta\,\slashed{\partial}\Lambda\psi - ie\xi\,\slashed{\partial}\Lambda\slashed{\partial}\rho$$
$$+ \tfrac{ie}{4}(1+2\beta+4\xi)\Box\Lambda\rho + ie(\alpha+\beta+2\xi)\partial_\mu\Lambda\partial^\mu\rho \quad (2.8)$$

This increases the number of parameters to 12! However, the requirement that the total action ($\mathcal{L} = \mathcal{L}_A + \mathcal{L}_\chi + \mathcal{L}_I$, where $\mathcal{L}_A$ is given in reference [9]) be invariant, mod $\Box^2\Lambda$, under these transformations reduces this number to five. It also makes it necessary to choose $2\zeta + \zeta' = 1$ which in turn makes $\mathcal{L}_\chi$ singular. Moreover, $A_\mu$ does not couple to a conserved current ($\partial_\mu J^\mu + \tfrac{1}{2}\Box J_+ \neq 0$) which implies violation of unitarity. The details are given in Appendix II.

These complications call for a deeper analysis with possible new reformulation which will be pursued in the next section.

## 3. Analysis and Conclusions

In this section we make a detailed analysis of the representation of the conformal group carried by the quantum spinor field and look for a possible way to characterize it. The notation in this section is explained in Appendix III.

The reproducing kernel for the degree $-2$ spinor has the following general form in Dirac's six-cone notation [5]:

$$\langle \chi(y)\bar{\chi}(y')\rangle = (y \cdot y')^{-2} + \tfrac{c}{2}\slashed{y}\slashed{y}'(y \cdot y')^{-3}. \tag{3.1}$$

This is the homogeneous propagator whose Fourier decomposition into positive energy states shows the modes that make up the quantum field. The physical modes contribute only to the first term, while the second term is made up of the gauge modes:

$$\chi = \slashed{y}\tau, \quad y \cdot \partial \tau = -3\tau.$$

The problem is to find a wave equation that characterizes (3.1). This propagator is uniquely characterized (modulo gauge) by homogeneity:

$$y \cdot \partial \chi = -2\chi. \tag{3.2}$$

However, this cannot be regarded as a wave equation, since its modification by interactions would imply an anomalous conformal degree, a property that may be in conflict with the concept of a gauge field. Moreover, it cannot be obtained by variation in a 4-dimensional action. This is because (3.2) implies that the spinor $\chi$ is not just a field over $x_\mu$, but it also depends on the homogeneity coordinate $x^+$ ($= y^5 + y^4$, see references [6] and [9]). Next, we turn to the Casimir invariants in the hope that they give the sought-after wave equation. The three Casimir operators give the equation $\slashed{y}\slashed{\partial}\chi = 0$, which is not satisfied by (3.1). In an irreducible representation, the Casimir operator is a multiple of the identity; however, in a non-decomposable representation, its difference from that is at worst nilpotent. Therefore, some positive integral power of the Casimir equation must be satisfied by (3.1). In the present case, the square of $\slashed{y}\slashed{\partial}\chi = 0$ is just (3.2). The analysis of (3.1) which is carried out in Appendix III shows that the "physical" and "gauge" modes satisfy:

$$\slashed{y}\slashed{\partial}P = 0, \quad \slashed{y}\slashed{\partial}G = 0,$$

respectively, while the "scalar" ($S$) satisfies:

$$\slashed{y}\slashed{\partial}S = 4\slashed{y}G.$$

Therefore, the condition $\slashed{y}\slashed{\partial}\chi = 0$ annihilates the scalar modes and being as such it is interpreted as the Lorentz condition for the spinor gauge module. Hence, the free action

$$\tfrac{1}{2}\int (dy)\bar{\chi}\slashed{y}\slashed{\partial}\chi, \quad (dy) = \delta(y^2)d^6y, \tag{3.3}$$

is not appropriate since the associated wave equation $\slashed{y}\slashed{\partial}\chi = 0$ is not satisfied by all the modes in (3.1).

There exists no invariant nontrivial wave operator of any integral degree (aside from (3.2)) that characterizes the representation propagated by (3.1). The same conclusion is reached when we look for an intertwining operator. For the most degenerate representations, like the one we are dealing with, it sometimes happens that elements of the enveloping algebra other than the Casimirs are fixed. Let

$$A_{\alpha\beta} = \tfrac{1}{4}\varepsilon_{\alpha\beta\gamma\delta\lambda\rho} L^{\gamma\delta} L^{\lambda\rho},$$

$$A_{\alpha\beta} = L_\alpha^\gamma L_{\gamma\beta} + L_\beta^\gamma L_{\gamma\alpha},$$

where $L_{\alpha\beta} \in so(4,2)$; then it may happen that

$$\left(A_{\alpha\beta} - a\Gamma_7 L_{\alpha\beta}\right)\chi = 0, \text{ and/or}$$

$$\left(B_{\alpha\beta} - b\eta_{\alpha\beta}\right)\chi = 0,$$

where the chirality projector $\Gamma_7 = \gamma_5 \oplus -\gamma_5$ and $a$, $b$ are constants. The results of such investigation, which is carried out in Appendix IV, can be stated in general terms: We construct the most general tensor operator $C_{\alpha_1 \ldots \alpha_n}$ for $n = 1, 2, \ldots$ under the requirements that it is intrinsic on the cone and annihilates (3.1). For $n = 1$, we also require that $\Gamma^\alpha C_\alpha = 0$ and $y^\alpha C_\alpha \equiv 0$ because no nontrivial invariant operator exists whose kernel contains (3.1). Thereafter, it is found that $C_\alpha$ must be trivial. So, we iterate this procedure for $n = 2, 3, \ldots$ and again we end up with the same trivial result for $n \leq 2$. By induction, we conclude that this representation is not characterized by covariant differential operators. This can be understood by noting that the generalized function $(y \cdot y')^{-2}$ in (3.1) is practically a $\delta$-distribution: $\int (dy)(y \cdot y')^{-2} f(y) \sim f(y')$. Therefore, its Fourier decomposition contains all possible gauge modes restricted only by homogeneity. Moreover, the corresponding distribution in Minkowski notation, given by (2.2) with $\lambda' = c$, has a Fourier transform whose inverse is not a finite homogeneous polynomial in the momentum.

With the failure of the foregoing attempt to characterize the whole spinor gauge module, we try the next best thing: Project out the physical sub-representation. In the notation of Section 2.1, this representation is carried by the Minkowski spinor component $\psi$ that satisfies the massless Dirac equation $\gamma^\mu \partial_\mu \psi = 0$. In Minkowski notation, the 8-spinor multiplet $\hat{\chi}(x) = \psi \oplus \sigma$ is related to $\chi(y)$ by

$$\hat{\chi}(x) = \left(y^5 + y^4\right)^2 \left[1 + \tfrac{1}{2}\left(\Gamma^5 + \Gamma^4\right)x^\mu \Gamma_\mu\right] \chi(y).$$

The operator that projects out $\psi$ is the invariant $\slashed{y}$. If we now let $\xi = \slashed{y}\chi$, then $\hat{\xi}(x) = 0 \oplus \psi$; $\partial^2 \xi = 0$ and $\slashed{y}\slashed{\partial}\chi = 0$ read the same in Minkowski notation: $\gamma^\mu \partial_\mu \psi = 0$.

Therefore, a physically equivalent theory written in terms of $\chi$ and $\xi$ can still be formulated with an action that is invariant under $\chi \to \chi + \slashed{y}\tau$ ($y \cdot \partial \tau = -3\tau$), and a boundary condition $\langle |\slashed{y}\slashed{\partial}\chi| \rangle = 0$. It is a perturbative quantum field theory defined by its propagators, vertices and Feynman rules. The propagators are those in (2.2) which are equivalent to (3.1) with $\lambda' = c$. The last two are highly singular and at first sight give the impression that the renormalization program may not be successful. However, it should be emphasized that "gauge invariance" ($\chi \to \chi + \slashed{y}\tau$) of the theory together with its conformal invariance take care of that. In other words, if we choose a covariant gauge in which $\lambda' = 0$ and observe, by conformal invariance, that $\psi$ and $\sigma$ couple to the vector potential only through the vertex $eA_+(\bar{\psi}\sigma + \bar{\sigma}\psi)$, then it is not difficult to see (*e.g.* by power counting) that the theory is "manifestly renormalizable". It is generally believed, however, that renormalization breaks conformal invariance and precisely for this reason the present program, with its unique gauge structure, should be pursued as far as possible. The 2-point function $\langle \xi(y)\bar{\xi}(y') \rangle = \slashed{y}\slashed{y}'(y \cdot y')^{-2}$ is analyzed in Appendix III and shown to contain only the physical massless modes.

Hence, a new direction is being followed from this point on. The free spinor action is written in terms of the gauge invariant "field strength" $\xi \equiv \slashed{y}\chi$, while the "spinor potential" $\chi$ plays the same role as the vector potential $A_\mu$ in electrodynamics [$F = dA$, where $d$ is the deRham cohomology operator which satisfies $d^2 = 0$, parallel to $\slashed{y}^2 = 0$]. The free action is

$$-\tfrac{1}{4}\int (dy)\bar{\xi}\,\partial^2 \xi$$

and on shell it is identical to (3.3). The free wave equation and boundary condition are

$$\partial^2 \xi = \slashed{y}\partial^2 \chi + 2\slashed{\partial}\chi = 0, \qquad \slashed{y}\xi = 0,$$

in respective analogy with

$$\partial^\mu F_{\mu\nu} = \Box A_\nu - \partial_\nu \partial A = 0, \qquad dF = 0.$$

The analogy with QED goes even further. The nontrivial propagator for the electromagnetic potential (with nonvanishing transverse part) does not satisfy Maxwell's equation. However, gauge fixing in QED eliminates this problem by giving a new wave equation that is compatible with the propagator. In the present case we were unable to find an analogous gauge fixing. [Gauge fixing in quantum electrodynamics modifies Maxwell's equation to $\Box A_\nu - \lambda \partial_\nu \partial A = 0$, $\lambda \neq 1$. The implication is that "spinor gauge fixing" should give the equation $\slashed{y}\partial^2\chi + 2\lambda\slashed{\partial}\chi = 0$, but this is "intrinsic" only for $\lambda = 1$. For a general homogeneous field $\chi$ of degree $-N$, the intrinsic wave equation is

$$\left[\slashed{y}\partial^2 + 2(N-1)\slashed{\partial}\right]\chi = 0,$$

and $\lambda = N - 1$. So in this language gauge fixing means an anomalous conformal degree for $\chi$.] Nevertheless, postulating (3.1) implies the existence of a reproducing kernel $K(y, y') \equiv \langle \chi(y)\bar{\chi}(y') \rangle$ that propagates the spinor field produced by a source $J$ as

$$\chi(y) = \int (dy') K(y, y') J(y')$$

and we stop short of interpreting $K$ as the inverse of a wave operator obtained from the kinetic part of a free action. Stated differently, we suppose that gauge fixing is performed in a nonconventional free Lagrangian field theory while a physically equivalent (same S-matrix) and gauge invariant (under $\chi \to \chi + \slashed{y}\tau$) theory is written in terms of $\chi$ and $\xi$. Another indication that the spinor $\xi$ has to be included in the formulation comes from conformal supersymmetry where the superfield is homogeneous of degree $-1$ [12].

In conclusion, we state that the problem of incompatibility of the wave equation and propagators which was raised in Section 2.1 is due to the gauge phenomenon of the spinor field. Only in this very special spinor representation, whose reproducing kernel is (3.1), do we encounter this. Any other representation which contains the physical component is fully reducible, hence, not as much interesting or special. This is a typical behavior of gauge theories and a sign of their singular character. The manifestation of this singular character is familiar in the case of the vector potential $A_\mu$. However, in the case of the "spinor potential" $\chi$, it takes the form of nonexistence of a differential operator whose kernel contains $\chi$. In this section we concluded that this does not pose any problem to the physics and gave a prescription to reformulate the theory in such a way that makes this conclusion justifiable.

The problem we are dealing with here is strikingly similar to that encountered in the formulation of singleton field theory in de Sitter space; precisely to its equivalent formulation on the cone ($y^2 = 0$) in $R^5$ with a metric $= \text{diag}(+ - - - +)$ [10]. The theory of the Di, the spinor singleton, is written in terms of a homogeneous spinor $\chi$ of degree $-3/2$ on the 5-cone and it is invariant under the "chiral gauge transformation".

$$\chi \to \chi + \slashed{y}\tau, \qquad y \cdot \partial \tau = -\tfrac{5}{2}\tau .$$

The free action and propagator are [10]:

$$\tfrac{1}{2}\int (dy)\bar{\chi}\slashed{y}\slashed{\partial}\chi, \quad (dy) = \delta(y^2) d^5 y$$

$$K(y, y') = (y \cdot y')^{-3/2}$$

but $\slashed{y}\slashed{\partial}K \neq 0$. However, the Di is given by $\xi = \slashed{y}\chi$ and propagated by $\slashed{y}\slashed{y}'(y \cdot y')^{-3/2}$. Supersymmetry incorporates the Rac, the scalar singleton, into the supermultiplet and off-shell closure is accomplished with the addition of one auxiliary scalar field. However, the treatment in reference [10] does not include the electromagnetic potential while preliminary results from unextended conformal supersymmetry does. Superconformal QED is described by a triplet contained in a homogeneous scalar superfield with degree of homogeneity $-1$.

This construction will not be carried out here but will be pursued in the near future as a project in its own right.

## Appendix I

The second order Casimir operator of the conformal algebra is:

$$C_2 = \tfrac{1}{2} L^{\mu\nu} L_{\mu\nu} + K_\mu P^\mu + 4iD - D^2$$

where $\{L_{\mu\nu}, P_\mu\}$ are the Poincare group generators and $D$, $K_\mu$ are, respectively, the generators of dilatation and special conformal transformations. $L_{\mu\nu}$ and $P_\mu$ act irreducibly and in the usual way on the spinor multiplet. The action of $K_\mu$ was given in (2.3) for $\alpha = \beta = \xi = 1$, while $D$ acts as follows:

$$D \begin{pmatrix} \rho \\ \psi \\ \sigma \end{pmatrix} = i \left[ x \cdot \partial + \frac{1}{2} \begin{pmatrix} 1 & 0 & 0 \\ 0 & 3 & 0 \\ 0 & 0 & 5 \end{pmatrix} \right] \begin{pmatrix} \rho \\ \psi \\ \sigma \end{pmatrix}.$$

Therefore, we get the following representation for $C_2$

$$C_2 \begin{pmatrix} \rho \\ \psi \\ \sigma \end{pmatrix}_{x=0} = - \begin{pmatrix} \tfrac{1}{4} & 0 & 0 \\ \not{\partial} & \tfrac{5}{4} & 0 \\ \Box & \not{\partial} & \tfrac{9}{4} \end{pmatrix} \begin{pmatrix} \rho \\ \psi \\ \sigma \end{pmatrix}_{x=0} .$$

This shows that the representation contains pieces that are not Weyl-equivalent [11] which is evidenced by the inequality of the diagonal elements of $C_2$.

The value of $C_2$ in the irreducible spinor representation $D(E_0, \tfrac{1}{2}, 0)$ of the conformal group is equal to $(E_0 - 2)^2 - \tfrac{5}{2}$. Therefore, the diagonal entries in $C_2$ show that the spinor multiplet contains the physical massless representation $D(\tfrac{3}{2}, \tfrac{1}{2}, 0)$ and its Weyl-equivalent $D(\tfrac{5}{2}, \tfrac{1}{2}, 0)$ corresponding to the value $-\tfrac{9}{4}$. However, it also contains the nonunitary $D(\tfrac{1}{2}, \tfrac{1}{2}, 0)$ associated with the diagonal element $-\tfrac{1}{4}$ in $C_2$. So, the whole representation is a direct sum of two pieces: a nondecomposable made up of the physical and "equivalent"; and a nonunitary. The latter is characterized by $(C_2 + \tfrac{1}{4})\chi = 0$ which is equivalent to $\psi = -\tfrac{1}{2}\not{\partial}\rho$ and $\sigma = -\tfrac{1}{4}\Box\rho$. The physical, on the other hand, is characterized by $(C_2 + \tfrac{9}{4})\chi = 0$ which gives $\rho = 0$ and $\not{\partial}\rho = 0$.

## Appendix II

The most general real six-current that is quadratic in the three spinors and whose components $(J_+, J_\mu, J_-)$ have conformal degrees (2,3,4) is:

$$J_+ = \tilde{a}\bar{\rho}\psi + \tilde{b}\bar{\rho}\vec{\partial}\rho + h.c.$$

$$J_\mu = \tfrac{1}{2}\bar{\psi}\gamma_\mu\psi + \hat{a}\bar{\rho}\gamma_\mu\sigma + \bar{\psi}\left(\hat{b}\vec{\partial}\gamma_\mu + \hat{c}\bar{\partial}_\mu + \hat{d}\gamma_\mu\vec{\partial} + \hat{f}\vec{\partial}_\mu\right)\rho$$
$$+\bar{\rho}\left(\tfrac{1}{2}\hat{g}\bar{\partial}\gamma_\mu\vec{\partial} + \tfrac{1}{2}\hat{h}\gamma_\mu\bar{\partial}\cdot\vec{\partial} + \hat{k}\gamma_\mu\Box + \hat{l}\bar{\partial}_\mu\vec{\partial} + \hat{m}\vec{\partial}_\mu\vec{\partial}\right)\rho + h.c.$$

$$J_- = a\bar{\psi}\sigma + b\bar{\rho}\vec{\partial}\sigma + c\bar{\rho}\vec{\partial}\sigma + d\bar{\psi}\vec{\partial}\psi$$
$$+\bar{\psi}\left(f\vec{\partial}\vec{\partial} + g\bar{\partial}\cdot\vec{\partial} + h\Box + k\vec{\Box}\right)\rho + \bar{\rho}\left(l\Box\vec{\partial} + m\vec{\partial}\Box + n\bar{\partial}\cdot\vec{\partial}\vec{\partial}\right)\rho + h.c.$$

Covariance under the action of special conformal transformation as given by (2.7) expresses the constant parameters $\tilde{a}, \tilde{b}, \hat{a}, ..., \hat{m}$ in terms of the 11 parameters $a,...,n$ of $J_-$. It also gives:

$$\tfrac{1}{4} - c + d + g - 2(f + k + n) + 4m = 0, \quad \text{and} \quad a = \tfrac{1}{2}.$$

Therefore, we end up with nine parameters. Now requiring that the total action be invariant under the gauge transformations (2.8) translates into:

$$\delta J_\pm^\mu = 0, \quad \text{and} \quad \delta\mathcal{L}_\chi = \partial^\mu \Lambda J_\mu - \tfrac{1}{2}\Box\Lambda J_+$$

which gives five more independent relations among the parameters of $J_-$ and requires that the constants in (2.8) satisfy:

$$\alpha = \tfrac{1}{2}, \quad \beta = -2\xi, \quad \text{and} \quad \xi\zeta = 0.$$

It also requires that the two parameters of $\mathcal{L}_\chi$ in (2.4) satisfy the singularity condition ($2\zeta + \zeta' = 1$). The wave equations obtained by variations in the full action do not give the current conservation equation

$$\partial_\mu J^\mu + \tfrac{1}{2}\Box J_+ = 0,$$

which was shown, in reference [6], as necessary for $A_+$ to be a generalized free field: $\Box^2 A_+ = 0$.

It may be worthwhile to mention that the five component conformal QED (where $J_+ = 0$) in this model has the following special case:

$$J_\mu = e\left(\bar{\psi} + \tfrac{1}{2}\bar{\rho}\vec{\partial}\right)\gamma_\mu\left(\psi + \tfrac{1}{2}\vec{\partial}\rho\right),$$

$$J_- = \tfrac{e}{2}\left(\bar{\psi} + \tfrac{1}{2}\bar{\rho}\vec{\partial}\right)\left(\sigma - \tfrac{1}{2}\vec{\partial}\psi\right) + h.c.$$

It satisfies the criterion for the J-B-A (Johnson-Baker-Adler) photon propagator, i.e. $\langle J_\mu(x) J_\nu(x') \rangle = 0$. It is also finite (verified up to three loops), however, the nontrivial sector of the theory is not unitary.

## Appendix III

Dirac's six-cone is Minkowski space compactified and embedded in $R^6$ as the surface $y^2 = y^\alpha y^\beta \eta_{\alpha\beta} = y^+ y^- - \vec{y}^2 = 0$ with the projection $\lambda y \simeq y$, for $\lambda \neq 0$. $\vec{y}$ stands for $(y^1, y^2, y^3, y^4)$ and $y^\pm = y^5 \pm i y^0$. Let $\{\Gamma_\alpha\}$ be a set of six $8 \times 8$ matrices satisfying the Clifford algebra $\{\Gamma_\alpha, \Gamma_\beta\} = 2\eta_{\alpha\beta}$ and as a basis we choose, ($\sigma^r$ are the three Pauli matrices):

$$\Gamma^\pm = \sigma^1 \pm i\sigma^2 = 2\sigma^\pm,$$

$$\vec{\Gamma} = \sigma^3 \vec{\gamma} = \left\{ \sigma^3 \begin{pmatrix} 0 & i \\ i & 0 \end{pmatrix}, \sigma^3 \begin{pmatrix} 0 & \sigma^r \\ -\sigma^r & 0 \end{pmatrix} \right\}.$$

To avoid cumbersome matrix indices we use an auxiliary 8-component spinor $\{Z_a\}$, $a = 1, 2, \ldots, 8$, and the spinor $\chi(y)$ is replaced by the scalar

$$\phi(y, Z) \equiv \bar{Z} \chi(y)$$

where $\bar{Z} = Z^\dagger \Gamma$, $\Gamma = \sigma^2 \gamma^5$, $\gamma^5 = \gamma^1 \gamma^2 \gamma^3 \gamma^4$, and $\Gamma_\alpha^\dagger = -\Gamma \Gamma_\alpha \Gamma$. The generators of so(4,2) algebra are

$$L_{\alpha\beta} = M_{\alpha\beta} + (\sigma_{\alpha\beta})_{ab} (\bar{Z}_a \bar{\partial}_b - Z_b \partial_a)$$

where

$$M_{\alpha\beta} = i(y_\alpha \partial_\beta - y_\beta \partial_\alpha), \quad \text{and} \quad \sigma_{\alpha\beta} = \frac{i}{4}[\Gamma_\alpha, \Gamma_\beta].$$

In this notation, the generators of the Cartan subalgebra are:

$$E \equiv L_{50} = y^- \partial_- - y^+ \partial_+ - \frac{1}{2} \begin{pmatrix} +1 & 0 \\ 0 & -1 \end{pmatrix}_{ab} (\bar{Z}_a \bar{\partial}_b - Z_b \partial_a),$$

$$L \equiv \frac{1}{2}(L_{12} + L_{34}) = \frac{1}{2}(M_{12} + M_{34}) + \frac{1}{2} \begin{pmatrix} 0 & \sigma^3 & & \\ \sigma^3 & 0 & & \\ & & 0 & \sigma^3 \\ & & \sigma^3 & 0 \end{pmatrix}_{ab} (\bar{Z}_a \bar{\partial}_b - Z_b \partial_a)$$

$$R \equiv \tfrac{1}{2}(L_{12} - L_{34}) = \tfrac{1}{2}(M_{12} - M_{34}) + \tfrac{1}{2}\begin{pmatrix} \sigma^3 & & 0 \\ & & \\ 0 & & \sigma^3 \\ & & 0 \end{pmatrix}_{ab} (\bar{Z}_a \bar{\partial}_b - Z_b \partial_a)$$

The energy raising and lowering operators are:

$$E_k^{\pm} \equiv L_{0k} \pm iL_{5k} = \pm(y^{\mp}\partial_k + 2y_k \partial_{\pm}) - (\sigma^{\pm}\gamma_k)_{ab}(\bar{Z}_a \bar{\partial}_b - Z_b \partial_a)$$

$k = 1,2,3,4$. Minimal weight irreducible representation of SO(4,2) are denoted by $D(E_0, j_1, j_2)$, where $E_0$ is the "conformal energy" and $j_1 - j_2$ is the helicity. If we write $Z$ as the sum of two 4-component spinors $Z_+ \oplus Z_-$, then the following table shows the weight of the coordinates $(y, Z)$ and the action of $E_i^{\pm}$ on them:

|  | $y^+$ | $y^-$ | $y_i$ | $Z_+$ | $Z_-$ | $\bar{Z}_+$ | $\bar{Z}_-$ |
|---|---|---|---|---|---|---|---|
| $(E_0, j_1, j_2)$ | $(-1,0,0)$ | $(+1,0,0)$ | $(0, \tfrac{1}{2}, \tfrac{1}{2})$ | $(\tfrac{1}{2})_{\pm}$ | $(-\tfrac{1}{2})_{\pm}$ | $(\tfrac{1}{2})_{\pm}$ | $(-\tfrac{1}{2})_{\pm}$ |
| $E_i^+$ | $2y_i$ | $0$ | $\delta_{ij} y^-$ | $0$ | $\gamma_i Z_+$ | $0$ | $\bar{Z}_+ \gamma_i$ |
| $E_i^-$ | $0$ | $-2y_i$ | $-\delta_{ij} y^+$ | $\gamma_i Z_-$ | $0$ | $\bar{Z}_- \gamma_i$ | $0$ |

where $(E_0)_{\pm} = D(E_0, \tfrac{1}{2}, 0) \oplus D(E_0, 0, \tfrac{1}{2})$ and $\bar{Z}_{\pm} = Z_{\pm}^{\dagger} \gamma^5$. The homogeneous two-point function is ($c$ is a gauge parameter):

$$K(yZ, y'Z') \equiv \langle \phi \phi'^* \rangle = \langle (\bar{Z}\chi)(\bar{\chi}Z)' \rangle = \bar{Z}Z'(2y \cdot y')^{-2} + c\bar{Z}\slashed{y}\slashed{y}'Z'(2y \cdot y')^{-3},$$

$$(2y \cdot y')^{-n} = e^n \left\{ 1 + 2ne(\vec{y} \cdot \vec{y}') + ne^2 \left[ -\vec{y}^2 \vec{y}'^2 + 2(n+1)(\vec{y} \cdot \vec{y}')^2 \right] \right.$$
$$\left. + 2n(n+1)e^3 (\vec{y} \cdot \vec{y}') \left[ -\vec{y}^2 \vec{y}'^2 + \tfrac{4}{3}(n+2)(\vec{y} \cdot \vec{y}')^2 \right] + ... \right\}$$

where $e = \left[ y^+(y'^+) \right]^{-1}$ and $y^-$ has been eliminated using $y^2 = 0$. The lowest energy level in $K$ is $3/2$ then it increases by integer steps:

$$K = K_{3/2} + K_{5/2} + K_{7/2} + ....$$

At every energy level $m$, we write $K_{m/2}$ as

$$K_{m/2} = \phi_p \phi_p'^* + \phi_s \phi_g'^* + \phi_g \phi_s'^* + \phi_g \phi_g'^*,$$

–51–

where $\phi_g = (\not{y})_{ab} \bar{Z}_a \bar{\partial}_b \tilde{\phi}_g$ is the gauge field, and every $\phi$ is written as a sum of irreducible representation of SO(4). Then:

$$K_{3/2} = ie^2 \bar{Z}_- Z'_+ = \phi_p \phi'^*_p,$$

where $\phi_p = i\bar{Z}_-/(y^+)^2$ is an absolute ground state for $D(3/2)_\pm$.

$$K_{5/2} = -i(1+c)e^2 \bar{Z}_+ Z'_- + 4ie^3(\vec{y}\cdot\vec{y}')\bar{Z}_- Z'_+ + ice^3 y^+ \bar{Z}_+(\vec{y}'\cdot\vec{\gamma})Z'_+$$
$$\quad - ice^3 y'^- \bar{Z}_-(\vec{y}\cdot\vec{\gamma})Z'_- + ice^3 \bar{Z}_-(\vec{y}\cdot\vec{\gamma})(\vec{y}'\cdot\vec{\gamma})Z'_+$$
$$\equiv \phi_p \phi'^*_p + \phi_s \phi'^*_g + \phi_g \phi'^*_s + \phi_g \phi'^*_g$$

where

$$\phi_p^{ij} = \frac{i}{\sqrt{3}} \frac{\bar{Z}_-}{(y^+)^3} \left( y^i \gamma^j + y^j \gamma^i - \tfrac{1}{2} \delta^{ij} \vec{y}\cdot\vec{\gamma} \right),$$

$$\phi_g = -i\frac{\bar{Z}_+}{(y^+)^2} - i\frac{\bar{Z}_- \vec{y}\cdot\vec{\gamma}}{(y^+)^3} = (\not{y})_{ab} \bar{Z}_a \bar{\partial}_b \left[ \frac{i\bar{Z}_-}{(y^+)^3} \right],$$

$$\phi_s = i\frac{\bar{Z}_+}{(y^+)^2} + \tfrac{c}{2} \phi_g.$$

$\phi_g$ is an absolute state for $D(5/2)_\pm$ while $\phi_s$ is a relative ground state for $D(5/2)_\pm$ and it is cyclic for the whole representation space. The leaking among these ground states can easily be verified:

$$E^i_- \phi_s = \phi_p \gamma^i, \qquad E^i_+ \phi_p = -\phi_g \gamma^i + ..., \qquad E^i_- \phi_p = E^i_- \phi_g = 0.$$

Therefore, we have established that $\chi$ carries the non-decomposable triplet (Gupta-Bleuler triplet):

$$D(5/2)_\pm \to D(3/2)_\pm \to D(5/2)_\pm$$

and since only these two representations are Weyl equivalent, we don't need to go to higher energy levels.

The ground states satisfy the following equations:

$$(\not{y}\not{\partial})_{ab} \bar{Z}_a \bar{\partial}_b \phi_p = (\not{y}\not{\partial})_{ab} \bar{Z}_a \bar{\partial}_b \phi_g = 0, \qquad (\not{y}\not{\partial})_{ab} \bar{Z}_a \bar{\partial}_b \phi_s = 4\phi_g.$$

Now doing the same analysis for

$$K = \left\langle (\bar{Z}\xi)(\bar{\xi}Z)' \right\rangle = \bar{Z}\not{y}\not{y}'Z'(2y\cdot y')^{-2},$$

–52–

we find that it describes the massless irreducible representation $D(3/2)_\pm$ with the following absolute ground state

$$\phi_p = i\frac{\bar{Z}_+}{y^+} + i\frac{\bar{Z}_- \vec{y}\cdot\vec{\gamma}}{(y^+)^2} = (\slashed{y})_{ab} \bar{Z}_a \bar{\partial}_b \left[\frac{-i\bar{Z}_-}{(y^+)^2}\right].$$

For the *Di* on the five-cone, the group is SO(3,2) and the homogeneous propagator [10] is:

$$\left\langle (\bar{Z}\chi)(\bar{\chi}Z)' \right\rangle = \bar{Z}Z'(y\cdot y')^{-3/2} + c\bar{Z}\slashed{y}\slashed{y}'Z'(y\cdot y')^{-5/2}.$$

Carrying out an analysis parallel to the one described above for SO(4,2), we find the triplet:

$$D(2,\tfrac{1}{2}) \to D(1,\tfrac{1}{2}) \to D(2,\tfrac{1}{2})$$

where $D(E_0,S)$ is the irreducible representation of SO(3,2) characterized by energy $E_0$ and spin $S$, and the gauge subspace $D(2,\tfrac{1}{2})$ is made up of modes all of the form

$$\chi = \slashed{y}\tau, \quad \text{and} \quad y\cdot\partial\tau = -\tfrac{5}{2}\tau$$

The physical sub-quotient, the *Di*, is carried by the "field strength" $\xi = \slashed{y}\chi$ in close analogy with the conformal spinor.

## Appendix IV

In this appendix we justify the conclusion of Section 3 that the representation carried by the "spinor gauge potential" is not characterized by covariant differential operators. We do that by attempting to construct such an operator under the requirements that it is intrinsic on the Dirac cone (when acting on $\chi$; $y\cdot\partial\chi = -2\chi$) and annihilates (3.1).

    1. Scalar operators: the intrinsic requirement gives the following complete list (*a* is a constant):

$$\slashed{y}, \quad \slashed{y}\slashed{\partial} + a, \quad \slashed{y}\partial^2 + 2\slashed{\partial}.$$

None of these kill (3.1).

    2. Vector operators ($C_\alpha$): the intrinsic requirement and $y^\alpha C_\alpha = \Gamma^\alpha C_\alpha = 0$ (from above) give the following complete list:

$$a\Gamma_\alpha \slashed{y} + y_\alpha(b\slashed{y}\slashed{\partial} - 6a), \quad \Gamma_\alpha \slashed{y}\slashed{\partial} - 3y_\alpha(b\slashed{y}\partial^2 + 2\slashed{\partial}),$$

where *a* and *b* are constants. Again none of these annihilate (3.1).

    3. Tensor operators ($S_{\alpha\beta} = S_{\beta\alpha}$, $A_{\alpha\beta} = -A_{\beta\alpha}$): the bases for $S_{\alpha\beta}$ are $\{\eta_{\alpha\beta}, y_\alpha y_\beta, \partial_\alpha\partial_\beta$ , $y_\alpha\Gamma_\beta + y_\beta\Gamma_\alpha, \Gamma_\alpha\partial_\beta + \Gamma_\beta\partial_\alpha\}$ and for $A_{\alpha\beta}$ they are $\{\Gamma_\alpha\Gamma_\beta - \Gamma_\beta\Gamma_\alpha, y_\alpha\partial_\beta - y_\beta\partial_\alpha,$

$y_\alpha \Gamma_\beta - y_\beta \Gamma_\alpha$, $\Gamma_\alpha \partial_\beta - \Gamma_\beta \partial_\alpha$, (no $\varepsilon_{\alpha\beta}{}^{\gamma\delta\lambda\rho}$, since its inclusion makes a chiral projection via $\Gamma_7 = \gamma_5 \oplus -\gamma_5$ which must vanish independently)}. The intrinsic requirement and $y^\alpha C_\alpha = \Gamma^\alpha C_\alpha = 0$ (from (2) above) leave no symmetric tensors that can kill (3.1). The same thing happens in the case of the antisymmetric tensor $A_{\alpha\beta}$.

By induction, we conclude that no such tensor of any finite rank annihilates (3.1).

**References to Section C**